\documentclass[twocolumn]{aastex631}

\usepackage{amsmath}
\usepackage{xspace}
\usepackage{xcolor}

\urldef{\morphVAC}\url{https://www.sdss.org/dr16/data_access/value-added-catalogs/}

\newcommand{\hkpc}{kpc/$h$\xspace}

\newcommand{\logOH}{$12 + \log(\text{O}/\text{H})$\xspace}
\newcommand{\HI}{\ion{H}{1}\xspace}
\newcommand{\MHI}{$M_\text{H{\sc i}}$\xspace}

\newcommand{\Mratio}{$M_\text{tot}/M_\text{vis}$\xspace}
\newcommand{\MratioS}{$M_\text{tot}/M_*$\xspace}
\newcommand{\MratioHI}{$M_\text{tot}/(M_* + M_\text{H{\sc i}})$\xspace}

\newcommand{\Ntot}{1988\xspace}
\newcommand{\NHI}{844\xspace}

\shorttitle{Mass ratio relationships}
\shortauthors{Douglass \& Demina}

\begin{document}

\title{Dependence of the ratio of total to visible mass on observable properties of SDSS MaNGA galaxies}

\correspondingauthor{Kelly A. Douglass}
\email{kellyadouglass@rochester.edu}

\author[0000-0002-9540-546X]{Kelly A. Douglass}
\affiliation{Department of Physics \& Astronomy, University of Rochester, 500 Wilson Blvd., Rochester, NY  14611}

\author[0000-0002-7852-167X]{Regina Demina}
\affiliation{Department of Physics \& Astronomy, University of Rochester, 500 Wilson Blvd., Rochester, NY  14611}

\begin{abstract}
  Using spectroscopic observations from the SDSS~MaNGA~DR15, we study the 
  relationships between the ratio of total to visible mass and various 
  parameters characterizing the evolution and environment of the galaxies in 
  this survey.  Measuring the rotation curve with the relative velocities of the 
  H$\alpha$ emission line across a galaxy's surface, we estimate each galaxy's 
  total mass.  We develop a statistical model to describe the observed 
  distribution in the ratio of total to visible mass, from which we extract a 
  galaxy's most probable value for this mass ratio.  We present the 
  relationships between the ratio of total to visible mass and several 
  characteristics describing galactic evolution, such as luminosity, gas-phase 
  metallicity, distance to the nearest neighbor, and position on the 
  color-magnitude diagram.  We find that faint galaxies with low metallicities, 
  typically in the blue cloud, have the highest ratios of total to visible mass.  
  This mass ratio is significantly reduced when we include the \HI mass in the 
  total visible mass, implying that feedback mechanisms are not as strong in 
  low-mass galaxies as previously thought.  Those galaxies that exhibit the 
  second highest ratios of total to visible mass are the brightest with high 
  metallicities, typically members of the red sequence or green valley.  AGN 
  activity is likely both the quenching mechanism and the feedback that drives 
  the mass ratio higher in these massive galaxies.  Finally, we introduce a 
  parametrization that predicts a galaxy's ratio of total to visible mass based 
  only on its photometry and luminosity.
\end{abstract}

\section{Introduction}

Current cosmological models indicate that the majority of matter in the universe 
is composed of dark matter \citep{Planck20}: material that interacts primarily 
through gravity.  Observational evidence for dark matter exists across most 
scales in the universe, from gravitational lensing of galaxy clusters 
\citep[and references therin]{Bartelmann10} to galaxy kinematics 
\citep[e.g.,][]{Freeman70,Bosma78,Carignan85,Salucci19}.  Dark matter 
simulations are able to reproduce the current distribution of galaxies 
\citep[e.g.,][]{Springel05}, indicating that dark matter's evolution throughout 
cosmic time dictates the formation and evolution of galaxies.

Understanding the quantity and distribution of dark matter in galaxies is 
crucial to our studies of galaxy formation and evolution.  To this end, we need 
to measure how much dark matter exists in galaxies, and we need to understand 
how the ratio of dark to visible matter influences a galaxy's star formation 
history.  Previous studies of the relationship between the ratio of either dark 
matter or total mass to visible mass or luminosity show that faint galaxies have 
the most dark matter relative to their stellar mass and that the brightest 
galaxies have more dark matter than those of intermediate luminosity 
\citep{Persic96,Strigari08,TorresFlores11,Karukes17,Ouellette17,Wechsler18,Behroozi19,DiPaolo19,Douglass19,AquinoOrtiz20}.  
In addition, \cite{Behroozi19} found that massive quenched galaxies reside in 
more massive halos than star-forming galaxies of the same stellar mass.


In this paper, we utilize the SDSS MaNGA survey \citep{MaNGA} to investigate the 
possible relationships of a galaxy's total, or dynamical, mass with its star 
formation history and local environment.  We use the relative velocities of the 
H$\alpha$ emission line across a galaxy's surface to measure the rotation curve 
of each galaxy, from which we estimate the galaxy's total mass, $M_\text{tot}$.  
The sum of the stellar mass, $M_*$, and the \HI mass estimates the total visible 
mass, $M_\text{vis}$, in each galaxy.  We present the correlation of \Mratio 
with various galaxy properties related to the galaxy's star formation history 
and environment.

\begin{figure}
    \centering
    \includegraphics[width=0.49\textwidth]{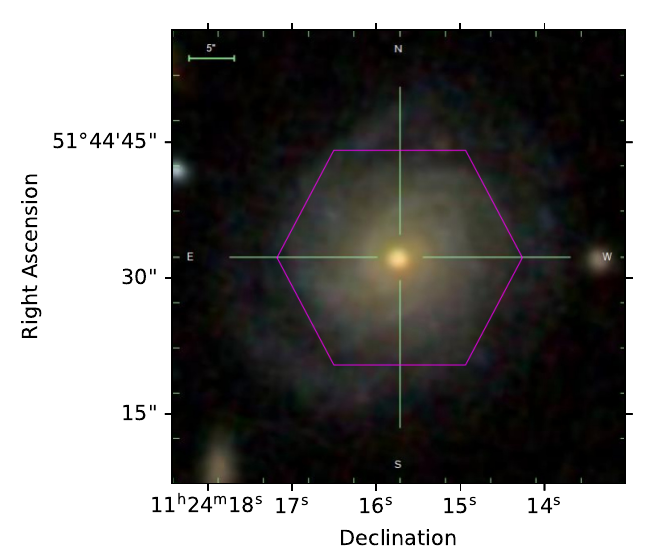}
    \caption{Example RGB composite image of a MaNGA galaxy (8997--9102) with the 
    IFU (magenta hexagon) overlayed \citep[made with the SDSS Marvin python 
    package by][]{SDSS-Marvin}.  As is typical with the MaNGA observations, the 
    IFU does not cover the entire visible component of the galaxy.}
    \label{fig:IFU}
\end{figure}

Statistical and systematic uncertainties on the measured rotational velocity 
introduce a significant amount of variation in the resulting ratios of total to 
stellar mass between the galaxies in the sample.  To adjust for this, we 
construct a statistical model that allows us to constrain the experimental 
uncertainties while extracting a more accurate estimate of the average \Mratio 
within each galaxy subsample.  In addition, as discussed in \cite{Douglass19} 
and visible in Figure \ref{fig:IFU}, the MaNGA data coverage is limited to the 
visible extent of the galaxy and to the size of the fiber-optic bundle (the 
integral field unit, or IFU).  Therefore, our measurement of the rotational 
velocity, $V(r)$, is limited to the covered region.  We extrapolate our 
parameterization of the fitted rotation curve to estimate $V(r)$ at the 90\% 
elliptical Petrosian radius, $R_{90}$, so that we study the same region of each 
galaxy.

Due to dark matter's lack of interaction via the electromagnetic force, there 
are limited statistics of the dark matter content of galaxies.  Studies have 
been restricted to measurements of weak gravitational lensing from galaxy 
stacking, which require assumptions about the galaxy's dark matter content 
\emph{a priori}, and to studies measuring the kinematics of a galaxy, which 
require multiple observations across the spatially resolved surface of the 
galaxy.  These kinematic studies are limited to only the closest objects and 
require a significant amount of observation time for each object.  To help 
alleviate these constraints, we introduce a parameterization of the ratio of 
total to stellar mass in a galaxy that requires galaxy photometry and redshift 
and can therefore be used on a much larger galaxy sample.

%
\section{SDSS MaNGA DR15 and Galaxy Selection}\label{sec:data}

We model the rotation curves of the velocity maps extracted from the H$\alpha$ 
emission line of galaxies in SDSS MaNGA DR15 \citep{SDSS15}.  MaNGA measures 
spectra across the face of each observed galaxy by placing a bundle of 
spectroscopic fibers (IFU) on each galaxy.  Each IFU contains between 19 and 127 
fibers, covering between 12" to 32" (corresponding to either $1.5R_e$ or 
$2.5R_e$\footnote{$R_e$ is the effective radius, or the 50\% Petrosian 
radius.}) of the surface of each galaxy \citep{Drory15}.  Two dual-channel 
spectrographs receive the light from the IFUs, covering a wavelength range of 
3600--10300\AA ~with a resolution of $\lambda/\Delta \lambda \sim 2000$.

At its conclusion, MaNGA will observe 10,000 nearby galaxies in the northern 
sky.  As described by \cite{Wake17}, targeted galaxies were selected to optimize 
the observed spatial resolution while maintaining a uniform distribution in 
luminosity.  The final galaxy selection is split into three components: the 
primary sample, for which each galaxy is observed out to $1.5R_e$, the secondary 
sample, where the surface coverage is extended out to $2.5R_e$, and the 
color-enhanced sample, augmenting the primary sample with low-mass red galaxies 
and high-mass blue galaxies.  The H$\alpha$ velocity map and $r$-band image 
processed by MaNGA's data analysis pipeline (DAP) are used to extract the 
galaxy's rotation curve.  The stellar mass density map processed by Pipe3D 
\citep{Sanchez16,Sanchez18} is used to extract the galaxy's stellar mass 
contained within the same region.  Absolute magnitudes are taken from the 
NASA-Sloan Atlas \citep[NSA;][]{Blanton11}.

We use the KIAS-VAGC \citep{Blanton05,Choi10} for photometric data (colors, 
color gradients, and inverse concentration indices) of the MaNGA galaxies.  We 
make use of the MPA-JHU value-added catalog\footnote{Available at 
\url{http://www.mpa-garching.mpg.de/SDSS/DR7}} for global emission line fluxes.  
Gas-phase metallicities are calculated using these flux ratios and the N2O2 
diagnostic described by \cite{Brown16}; this method has an systematic 
$\sim$0.1~dex uncertainty in the calculated metallicity.

All of these value-added catalogs (NSA, KIAS-VAGC, MPA-JHU) are based on the 
SDSS Data Release 7 \citep[DR7;][]{SDSS7} galaxy sample, which we use for 
finding a MaNGA galaxy's nearest neighbor.  SDSS DR7 employed a drift scanning 
technique to conduct a wide-field multiband imaging and spectroscopic survey 
that covered approximately one quarter of the northern sky.  A dedicated 2.5-m 
telescope at the Apache Point Observatory in New Mexico was used to take the 
photometric data in the five-band SDSS system: $u$, $g$, $r$, $i$, and $z$ 
\citep{Fukugita96,Gunn98}.  Galaxies with a Petrosian $r$-band magnitude 
$m_r < 17.77$ were chosen for follow-up spectroscopic analysis using two double 
fiber-fed spectrometers and fiber plug plates with a minimum fiber separation 
of 55" \citep{Lupton01,Strauss02}.  The wavelength coverage of the spectrometers 
used in this first stage of SDSS had an observed wavelength range of 
3800--9200\AA ~with a resolution of $\lambda/\Delta \lambda \sim 1800$ 
\citep{Blanton03}.

\HI mass estimates are from the \HI--MaNGA Data Release 2 \citep{Stark21} and 
the final Arecibo Legacy Fast ALFA (ALFALFA) data release \citep{Haynes18}.  
\HI--MaNGA is designed to conduct follow-up observations of all MaNGA galaxy 
targets on the Robert C. Byrd Green Bank Telescope (GBT) in Green Bank, West 
Virginia.  The second data release of this program contains observations of 
2135 MaNGA galaxies.  To supplement the GBT observations, \cite{Stark21} also 
include a cross match of the MaNGA DR15 sample with ALFALFA, a blind, 2-pass 
drift survey detecting extragalactic \HI that was conducted at the Arecibo 
Observatory in Arecibo, Puerto Rico.  ALFALFA adds \HI detections for 1021 MaNGA 
DR15 galaxies.  There are a total of \NHI galaxies with \HI detections in our 
final galaxy sample.

\subsection{Color-magnitude classification}

\begin{figure}
    \includegraphics[width=0.45\textwidth]{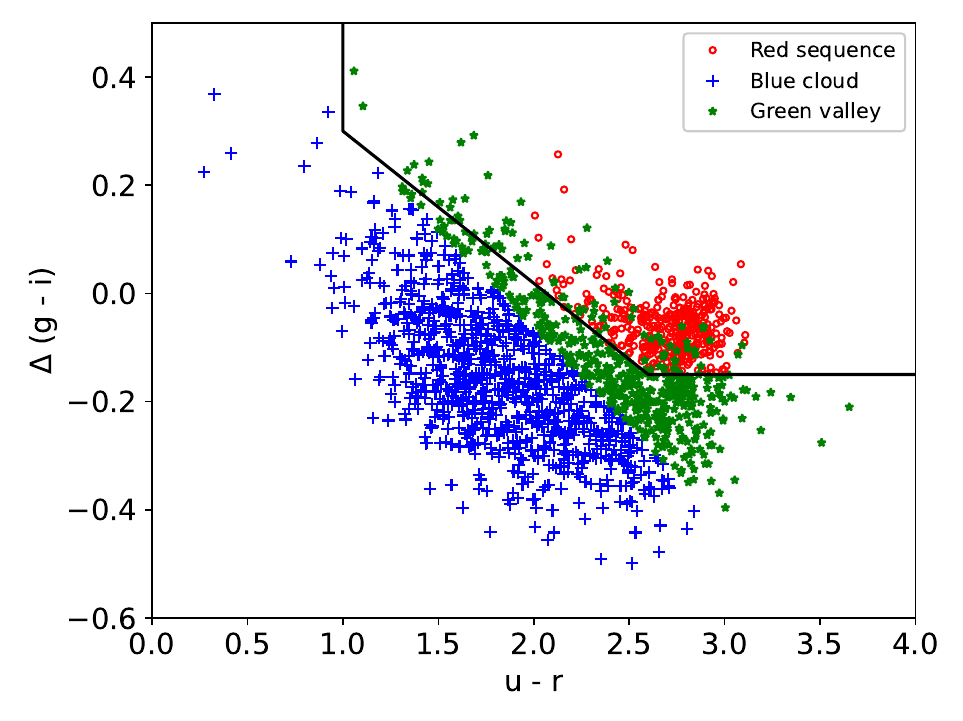}
    \caption{$\Delta (g - i)$ color gradient versus  $u - r$ color for our 
    sample of SDSS MaNGA galaxies, marked by their color-magnitude diagram 
    classification: open red circles for the red sequence, green stars for the 
    green valley, and blue crosses for the blue cloud.  The black boundary is 
    the separation between early- and late-type galaxies as defined by 
    \cite{Park05}.}
    \label{fig:CMD_class}
\end{figure}

To study the relationship between the mass ratio and the evolutionary stage of a 
galaxy, we separate the galaxies into populations in the color-magnitude 
diagram.  Fainter, bluer galaxies generally belong to the blue cloud, while 
brighter, redder galaxies belong to the red sequence.  Galaxies are theorized to 
transition between these two populations through the green valley 
\citep{Martin07}.  Our classification of the galaxies into one of these three 
populations is a modification of the photometric classification scheme described 
in \cite{Douglass_thesis}, which is based on the morphological classifications 
of \cite{Park05} and \cite{Choi10}.  Galaxies are classified based on their 
inverse concentration index, $c_\text{inv}$, and their position in the color 
($u - r$) -- color gradient ($\Delta (g - i)$) plane.  We define a phase angle 
$\theta$ such that
\begin{equation}
    \theta = \tan^{-1} \left( \frac{-\Delta (g - i) + 0.3}{(u - r) - 1} \right).
\end{equation}
We then classify the galaxy in either the blue cloud, green valley, or red 
sequence based on the following criteria:
\begin{description}

  \item[Red sequence] Normal early-type galaxies (galaxies above the boundary 
  defined  by the points (3.5, -0.15), (2.6, -0.15), and (1.0, 0.3) in the 
  $u - r$ versus $\Delta (g - i)$ space with $u - r > 2$ and 
  $c_\text{inv} \lesssim 3.8$).
  
  \item[Green valley] Blue early-type galaxies (galaxies above the boundary 
  defined in the red sequence description with $u - r < 2$); Normal late-type 
  galaxies that would otherwise be normal early-type galaxies except for their 
  high $c_\text{inv}$; Normal late-type galaxies (galaxies below the boundary 
  defined in the red sequence description) with $\theta < 20^\circ$.

  \item[Blue cloud] Normal late-type galaxies (galaxies below the boundary 
  defined in the red sequence description) with $\theta > 20^\circ$.
  
\end{description}
Figure \ref{fig:CMD_class} shows the results of this classification on our 
sample of SDSS MaNGA galaxies.  Since we require our objects to be dominated by 
rotational motion as described below in Section~\ref{sec:selection}, we would 
expect very few galaxies in our sample to be in the red sequence.  However, we 
find a significant number of these objects in our final galaxy sample.  Upon 
visual inspection, these galaxies appear to be a mixture of both red disk 
galaxies with little to no star formation and elliptical galaxies still 
supported by rotation.

\subsection{Modeling the velocity map}

\begin{figure*}
    \includegraphics[width=0.24\textwidth]{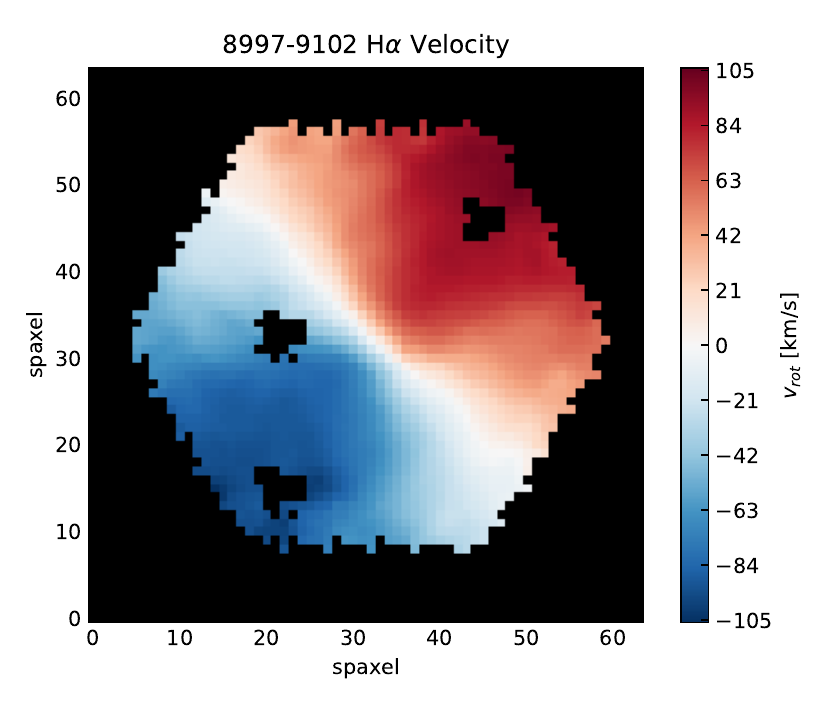}
    \includegraphics[width=0.24\textwidth]{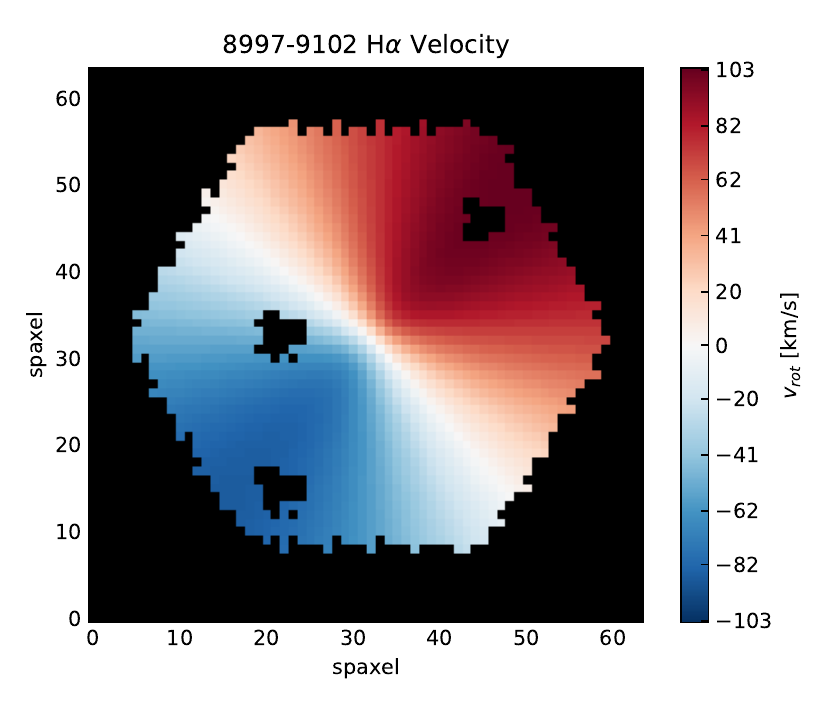}
    \includegraphics[width=0.24\textwidth]{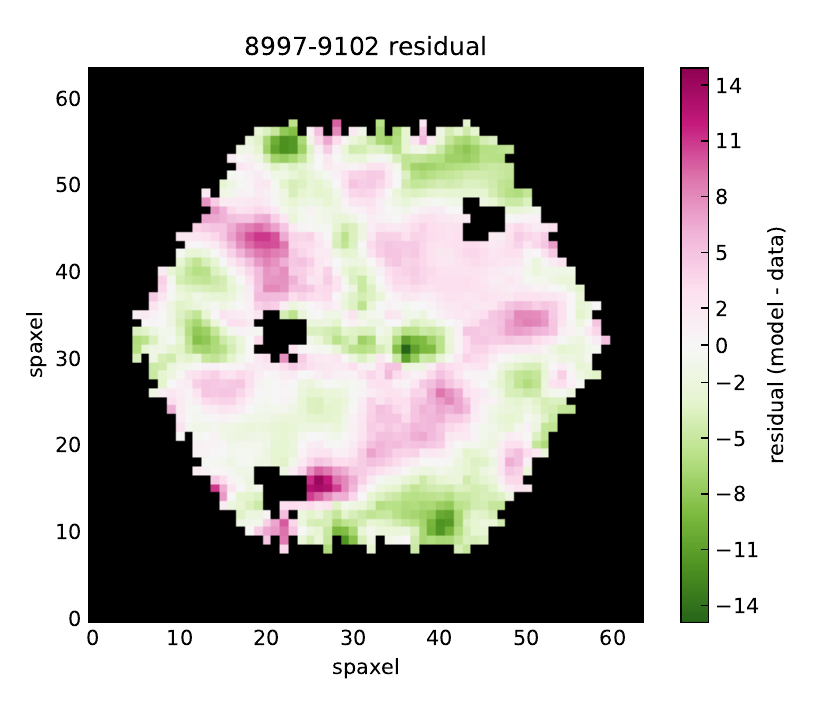}
    \includegraphics[width=0.24\textwidth]{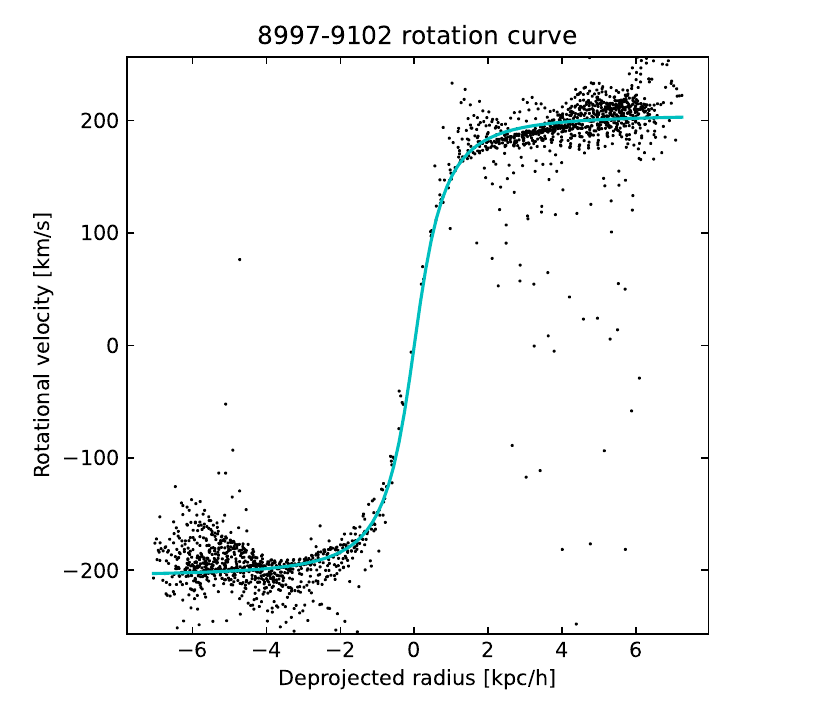}
    \includegraphics[width=0.24\textwidth]{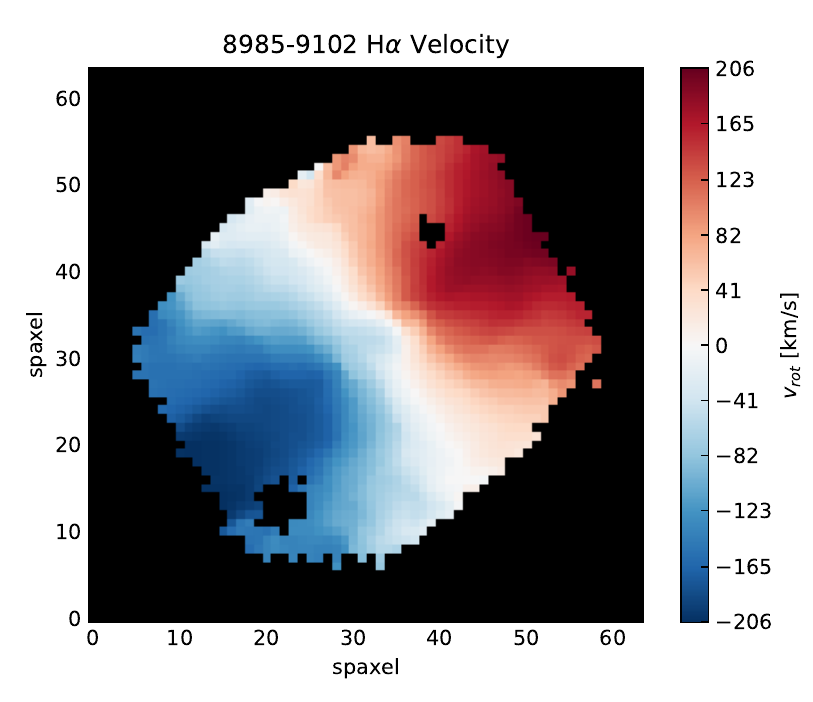}
    \includegraphics[width=0.24\textwidth]{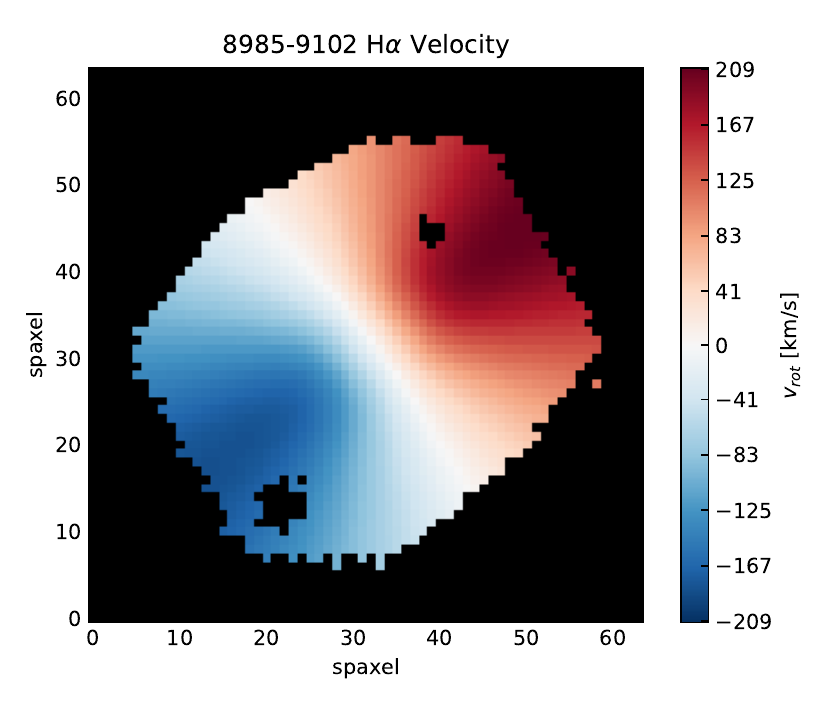}
    \includegraphics[width=0.24\textwidth]{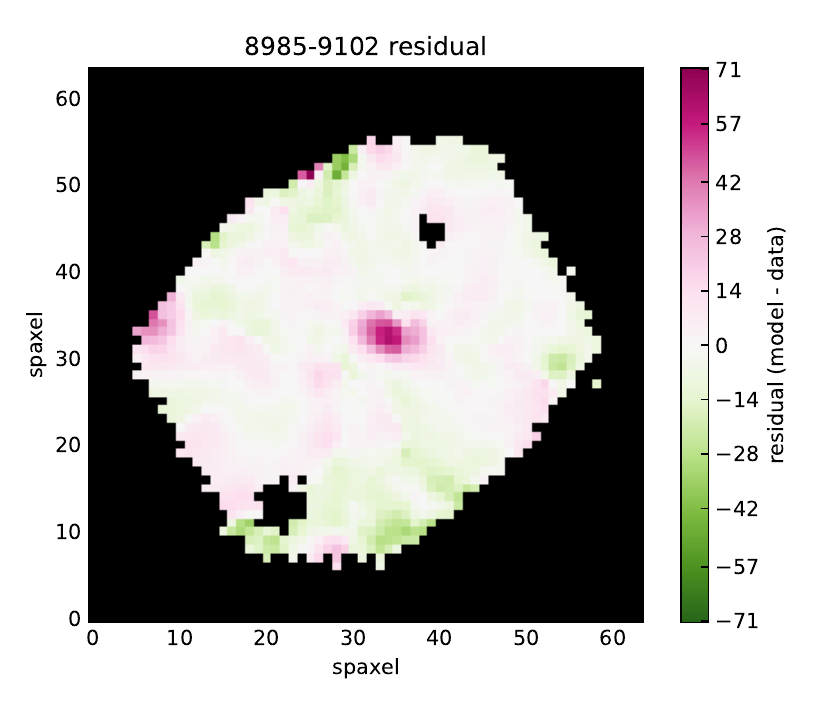}
    \includegraphics[width=0.24\textwidth]{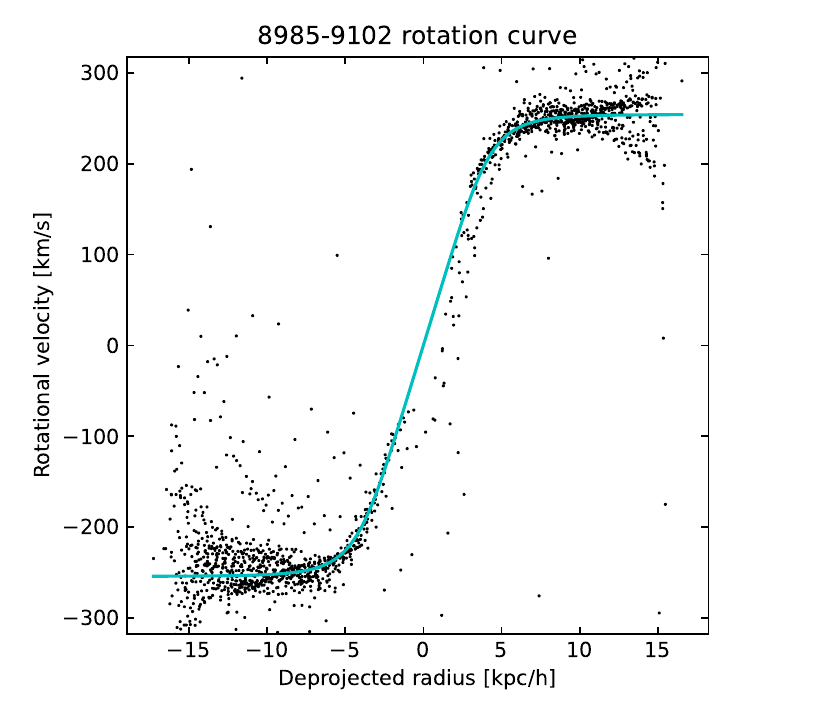}
    \caption{An example of the H$\alpha$ velocity map from the MaNGA DAP (first 
    column), our best-fit model of the field (second column), the residual 
    between our best-fit model and the data (third column), and the deprojected 
    rotation curve (fourth column) for an unbarred spiral galaxy (Sc, top row) 
    and barred spiral galaxy (SBbc, bottom row).}
    \label{fig:model_map}
\end{figure*}

We use a galaxy's H$\alpha$ velocity map to estimate its total dynamical mass.  
As provided by the SDSS MaNGA DAP, spaxels must have a data quality bit of 0 to 
be included in the analysis.  Galaxies are also required to have velocity maps 
with a smooth gradient.  See \cite{Douglass19} for a more detailed explanation 
of this velocity gradient measurement.

Each velocity field is fit with the parameterization 
\citep{BarreraBallesteros18}:
\begin{equation}\label{eqn:rot_curve}
    V(r) = \frac{V_{\text{max}} r}{(R^\alpha_{\text{turn}} + r^\alpha)^{1/\alpha}}
\end{equation}
where $V(r)$ is the tangential velocity at a given deprojected radius $r$, 
$V_{\text{max}}$ is the magnitude of the velocity at the rotation curve's 
plateau, $R_{\text{turn}}$ is the deprojected radius where the curve transitions 
from increasing to flat, and $\alpha$ is the degree of sharpness of the rotation 
curve.  The free parameters in the fit are $V_{\text{max}}$, $R_{\text{turn}}$, 
and $\alpha$.  We use this parameterization to determine the maximum rotational 
velocity for each galaxy, from which we can calculate the galaxy's total mass.  
Future work will include a decomposition of the rotation curve into the various 
mass components (bulge, disk, halo) to study the structure of the dark matter 
halo in more detail.  All distances are measured in units of Mpc/$h$, where the 
reduced Hubble constant $h$ is defined by $H_0 = 100h$~km/s/Mpc.

In addition to the three free parameters described above, we also fit for a 
galaxy's systemic velocity, kinematic center, inclination angle, and rotation 
angle on the sky.  We iterate through different combinations of these parameters 
to find the model velocity map that best represents the data.  See 
Figure~\ref{fig:model_map} for example H$\alpha$ velocity maps and best-fit 
models.

We only model the rotation component of the velocity maps with this 
parameterization.  It has been shown 
\citep{Valenzuela07,Randriamampandry15,Oman19} that rotational velocities can be 
over/underestimated when non-circular motions are not modeled in the central 
region of a galaxy, resulting from the presence of e.g., a bar.  In this study, 
we are only concerned with estimating the maximum rotational velocity of the 
outer extent of the galaxy, where non-circular motion is negligible.  
Figure~\ref{fig:model_map} shows both an unbarred (top row) and barred (bottom 
row) spiral galaxy.  It is readily apparent that our parameterization of the 
velocity map successfully reproduces the rotational velocity component 
irrespective of morphological type.

We define the best model velocity map based on which model produces the smallest 
$\chi^2_\nu$, where the fit statistic $\chi^2_\nu$ is normalized by the 
difference of the number of unmasked spaxels and the number of degrees of 
freedom of the fit (eight).  We find this best-fit model map based on four 
different data masks:
\begin{description}
    \item[Default] (spaxels with a data quality bit $> 0$ are masked)  We find a 
    best-fit model using this mask via two different criteria:
    \begin{itemize}
        \item The model with the smallest 
        $\chi^2_\nu = \sum ((\text{data} - \text{model})/\text{uncertainty})^2$
        
        \item The model with the smallest residual, 
        $\sum (\text{data} - \text{model})^2$
    \end{itemize}
    
    \item[Continuous] (helps to mask foreground artifacts)  We first bin all of 
    the unmasked spaxels with a bin width of 10~km/s.  Starting with the bin 
    containing the most spaxels, we define the velocity bounds based on the 
    first empty bin found in both directions from this central bin.  All spaxels 
    with velocities outside of this range are masked.  We then find the model 
    map that has the smallest $\chi^2_\nu$ using this continuous mask.
    
    \item[High S/N] (removes spaxels with low S/N H$\alpha$ emission)  We mask 
    all spaxels with a S/N in the H$\alpha$ flux less than 5 and find the model 
    which minimizes $\chi^2_\nu$.
    
    \item[No AGN] (helps to mask spaxels that likely contain emission from AGN, 
    defined as bins with unusually high velocity dispersion) We first bin all 
    unmasked spaxels in the H$\alpha$ velocity dispersion map with a bin width 
    of 10~km/s.  Starting with the bin containing 0~km/s, we define the velocity 
    dispersion bounds based on the first empty bin found in both directions from 
    this central bin.  All spaxels with velocities outside of this range are 
    masked.  We then find the model map that has the smallest $\chi^2_\nu$ using 
    this non-AGN mask.
\end{description}
Of these five best-fit models, we select the map with the smallest $\chi^2_\nu$ 
of those models with $\alpha < 100$.

\begin{figure}
    \includegraphics[width=0.47\textwidth]{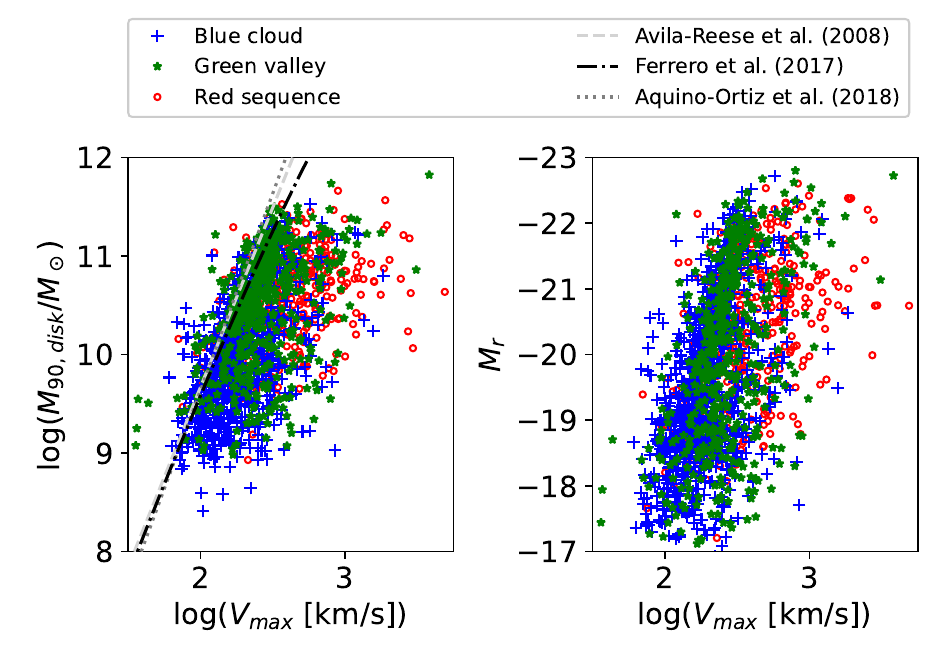}
    \caption{The Tully-Fisher relation (TFR) ---  the disk mass (left) and 
    absolute magnitude in the SDSS $r$-band, $M_r$ (right) as a function of the 
    logarithm of the maximum velocity --- for blue cloud (blue crosses), green 
    valley (green stars), and red sequence (red circles) galaxies.  For 
    comparison, the linear relationships of the baryonic TFR derived in 
    \cite{AvilaReese08, Ferrero17} and \cite{AquinoOrtiz18} are shown on the 
    left.}
    \label{fig:TF}
\end{figure}

To confirm that our best-fit models result in realistic galaxy kinematics, we 
plot the distribution of fitted maximum velocities as a function of the 
luminosity, $M_r$, and disk mass 
\citep[the Tully-Fisher relation, TFR;][]{Tully77} in Figure~\ref{fig:TF}.  As 
expected, we find a strong positive correlation between the luminosity and 
maximum velocity.  We also compare our baryonic TFR with the linear fits derived 
by \cite{AvilaReese08, Ferrero17} and \cite{AquinoOrtiz18}.  As the left-hand 
plot in Figure~\ref{fig:TF} shows, our velocities agree with their results.

\subsection{Modeling the stellar mass}

\begin{figure*}
    \includegraphics[width=0.49\textwidth]{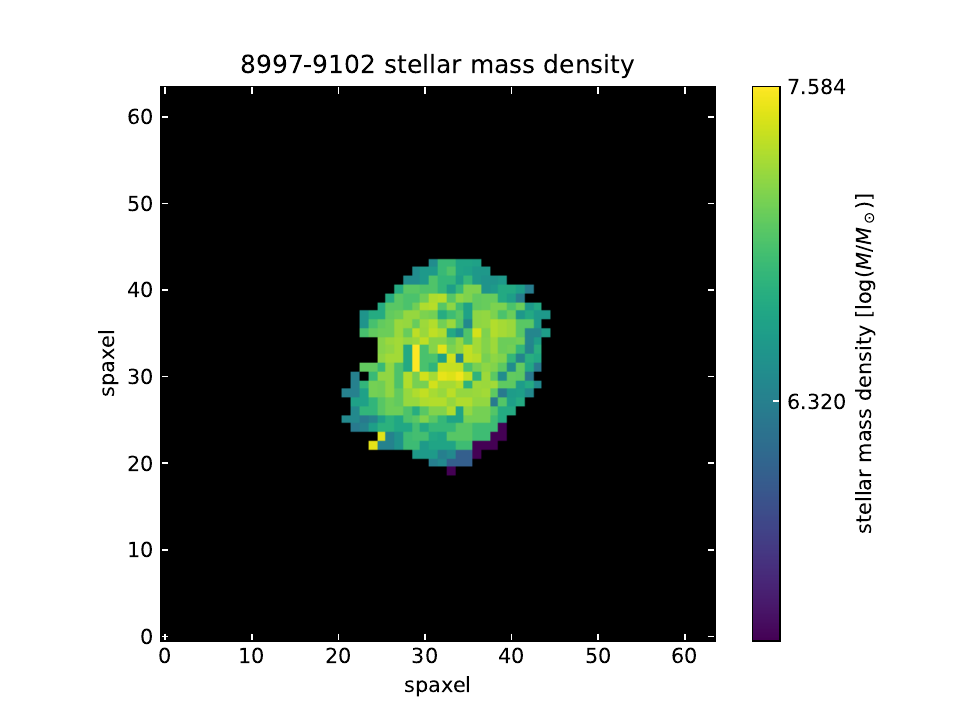}
    \includegraphics[width=0.49\textwidth]{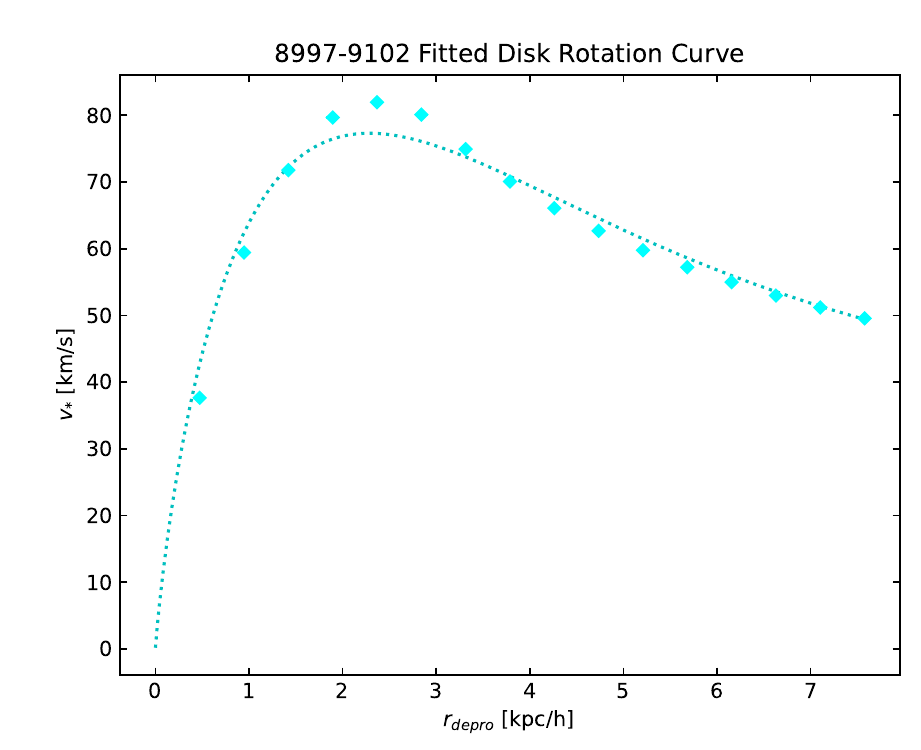}
    \caption{An example of the stellar mass density map from the Pipe3D analysis 
    pipeline (left) and our best fit to the rotation curve extracted from this 
    map (right).}
    \label{fig:smass_fit}
\end{figure*}

\begin{figure}
    \centering
    \includegraphics[width=0.49\textwidth]{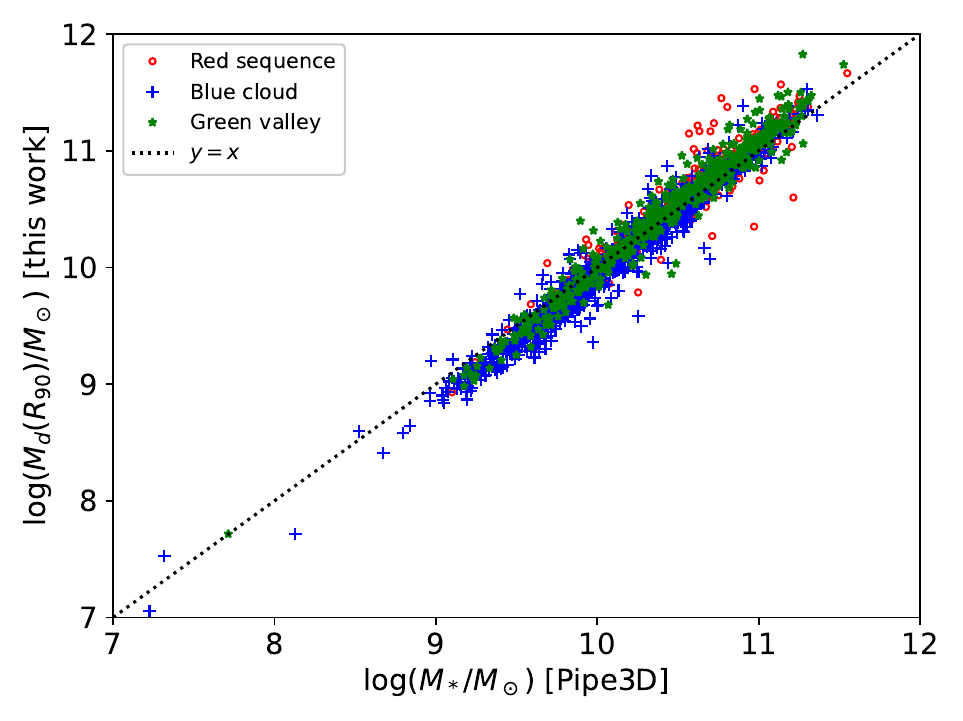}
    \caption{Comparison between our best-fit estimate of the disk mass within 
    $R_{90}$, $M_d (R_{90})$, and the stellar mass, $M_*$, as derived by Pipe3D, 
    for galaxies in the blue cloud (blue crosses), green valley (green stars), 
    and red sequence (red circles).  The black dotted line represents equality 
    between the two quantities.  There is good agreement between the two sets of 
    measurements.}
    \label{fig:Mstar_compare}
\end{figure}

To estimate the total stellar mass within the same radius for which we measure 
the total mass, we fit for the rotation curve due only to the stellar component 
of each galaxy.  We use the kinematic center, inclination angle, and position 
angle on the sky from the best-fit model velocity map to define the ellipse 
which corresponds to a given orbital radius in the galaxy.  Increasing the 
orbital radius in step sizes of 2~spaxels, we compute the sum of the stellar 
mass density per spaxel within these concentric ellipses of the stellar mass 
density map provided by the Pipe3D MaNGA analysis pipeline 
\citep{Sanchez16,Sanchez18}.  An example of this stellar mass density map is 
shown on the left in Figure \ref{fig:smass_fit}.  With this, we have the stellar 
mass as a discretized function of radius, $M_* (r)$.

Assuming that the stellar mass is the primary component of the galaxy's disk, we 
model the extracted stellar mass rotation curve as an exponential disk 
\citep[a thin disk without perturbation;][]{Freeman70}, where
\begin{equation}\label{eqn:V_disk}
    V_*(r)^2 = 4\pi G\Sigma_d R_d y^2 [I_0 (y) K_0 (y) - I_1 (y) K_1 (y)].
\end{equation}
Here, $V_* (r)$ is the rotational velocity due to the disk component of the 
galaxy, $\Sigma_d$ is the central surface mass density of the disk, $R_d$ is the 
scale radius of the disk, $y = r/2R_d$, and $I_i$ and $K_i$ are the modified 
Bessel functions \citep{Sofue13}.  The free parameters in this fit are 
$\Sigma_d$ and $R_d$.  The total disk mass within some radius $r$ is then
\begin{align}
    M_d (r) &= 2\pi \Sigma_d \int_0^r r e^{-r/R_d} \, dr \\
            &= 2\pi \Sigma_d R_d \left[ R_d - e^{-r/R_d} (r + R_d) \right]. \label{eqn:M_disk}
\end{align}
An example of the best fit to this stellar mass rotation curve is shown on the 
right in Figure \ref{fig:smass_fit}.  We compare our estimates of the disk mass 
within $R_{90}$ to that derived by the Pipe3D analysis pipeline.  As seen in 
Figure~\ref{fig:Mstar_compare}, the results of our fitting to the stellar mass 
density maps of Pipe3D are in good agreement with the Pipe3D estimates of the 
total stellar mass in the system.

\subsection{Galaxy selection criteria}\label{sec:selection}

\begin{figure}
    \centering
    \includegraphics[width=0.49\textwidth]{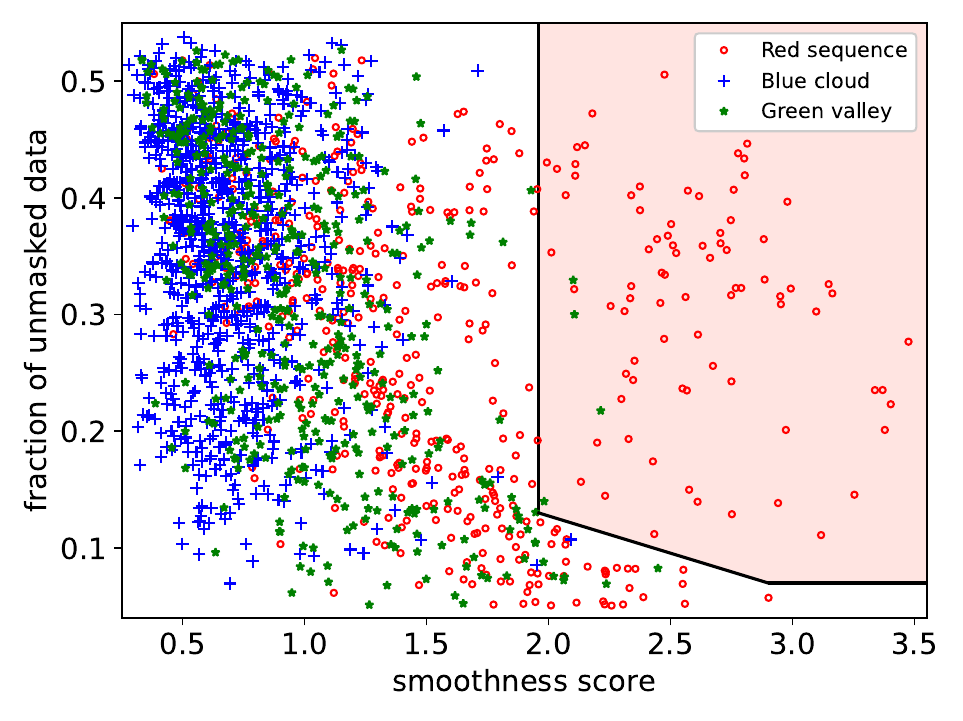}
    \caption{Relationship between the smoothness score and fraction of the 
    velocity maps that are unmasked for those galaxies in our sample.  
    Determined via visual inspection, objects above the black boundary (within 
    the red shaded region) have velocity maps which do not exhibit rotational 
    motion and are therefore excluded from our study.}
    \label{fig:elliptical_galaxy_cut}
\end{figure}

\begin{figure}
    \includegraphics[width=0.5\textwidth]{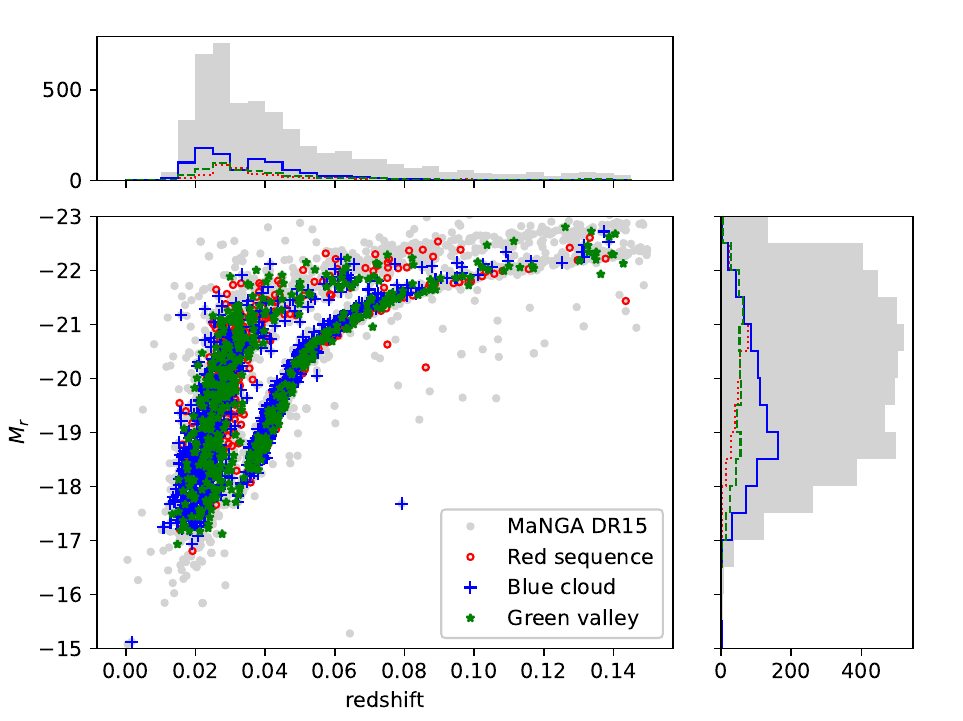}
    \caption{Distribution of our galaxy sample over luminosity and redshift.  
    Galaxies classified as part of the blue cloud are shown as blue crosses, 
    green valley galaxies are depicted as green stars, and red sequence objects 
    are shown as open red circles.  The corresponding distributions for each 
    property are shown in the histograms, with the same colors as defined above.  
    The entire SDSS MaNGA DR15 sample is shown in light gray.}
    \label{fig:sample_stats}
\end{figure}

After fitting both the H$\alpha$ velocity map and the stellar mass rotation 
curve, we restrict our sample to only include objects with successful fits for 
both of these models.  We therefore reject any fit with
\begin{itemize}
    \item $\alpha > 99$
    \item An inclination angle $> 86.4^\circ$
    \item Velocity maps with more than 95\% of their data masked
\end{itemize}
The first two requirements remove fits with an unsuccessful minimization of 
$\chi^2_\nu$ and therefore an untrustworthy and/or nonphysical model.  To ensure 
that there are sufficient unmasked spaxels to fit, visual inspection revealed 
that no more than 95\% of the total velocity map can be masked (this includes 
those data points outside the IFU footprint; for comparison, a completely 
unmasked IFU field corresponds to $\sim$55\% of the velocity map array 
unmasked).

Visual inspection of some of the stellar mass density maps produced by Pipe3D 
revealed that some maps have very sparse data and/or unrealistic stellar surface 
mass densities.  Fitting to these maps results in an unexpectedly low estimate 
for $M_d (R_{90})$, and therefore an extremely high ratio of the total to disk 
mass, $M_\text{tot}/M_d (R_{90})$.  We therefore also require that 
$M_\text{tot}/M_d (R_{90}) < 1050$.

We are focusing only on disc galaxies in this study (galaxies which are 
supported by rotation).  Requiring a successful model velocity map by 
implementing the above criteria removes most elliptical galaxies, mergers, and 
interacting systems.  We further remove all objects which have been determined 
to have some evidence of a disrupted velocity field.  Using the visual 
morphological classifications from the MaNGA Visual Morphologies from SDSS and 
DESI images catalog\footnote{Available at\\ \morphVAC}
and the morphological classifications from \cite{DominguezSanchez18}, we remove 
all objects with evidence of tidal debris (from the former) and a probability 
greater than 0.97 that it is a merger (from the latter).
To ensure that there are no spurious elliptical galaxies that remain in our 
sample, we also remove all objects which lie above the boundary shown in 
Figure~\ref{fig:elliptical_galaxy_cut}, defined by the points (1.96, 0.13) and 
(2.9, 0.07).  Visual inspection of these objects reveal that their velocity 
fields are dominated by random motion instead of rotation.

The proceeding analysis is conducted on a sample composed of disc galaxies.  
Since most elliptical galaxies are extremely bright, this introduces a selection 
bias towards fainter magnitudes.  Our velocity parameterization of 
Eqn.~\ref{eqn:rot_curve} assumes a flat rotation curve, which therefore 
preferentially eliminates galaxies with low dark matter content (for which a 
rotation curve is dominated by the bulge's mass).  Of the 4815 galaxies with IFU 
spectra available in the Pipe3D analysis of the SDSS DR15 MaNGA survey, we 
successfully model the velocity map for \Ntot galaxies.  The distribution over 
luminosity and redshift for our final sample of galaxies (relative to the full 
SDSS MaNGA DR15) is shown in Figure~\ref{fig:sample_stats}.

\section{Estimating the mass components}

\subsection{Total mass}

\begin{figure*}
    \centering
    \includegraphics[width=0.48\textwidth]{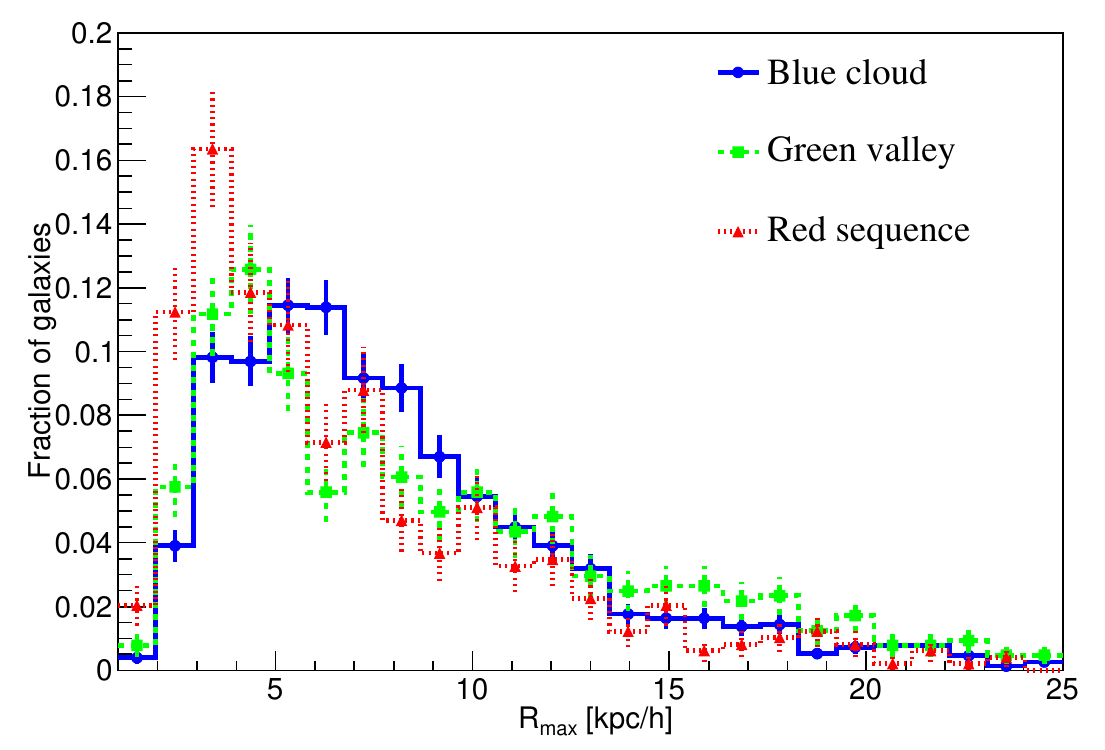}
    \includegraphics[width=0.48\textwidth]{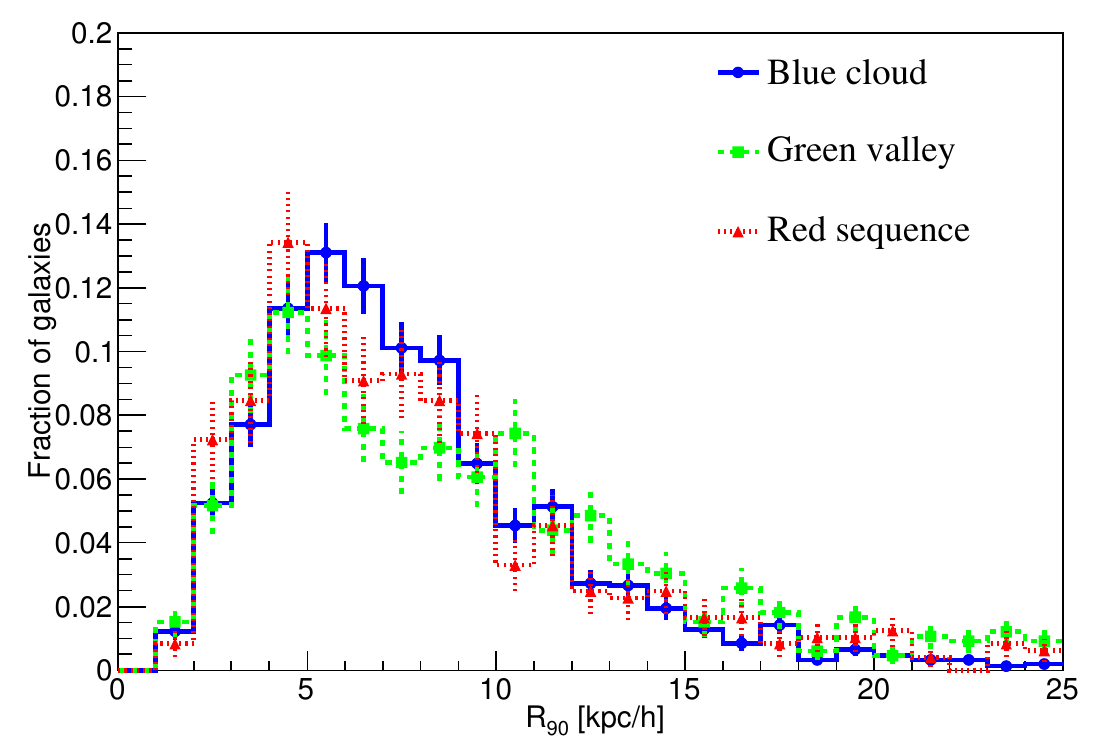}
    \caption{The distribution over the maximum radius to which the data extends 
    (left) and 90\% elliptical Petrosian light radius (right) to which the 
    rotational curves are evaluated for blue cloud (blue solid line), green 
    valley (green dashed line), and red sequence (red dotted line) galaxies.}
    \label{fig:Fr_Rmax}
\end{figure*}

\begin{figure*}
    \centering
    \includegraphics[width=0.5\textwidth]{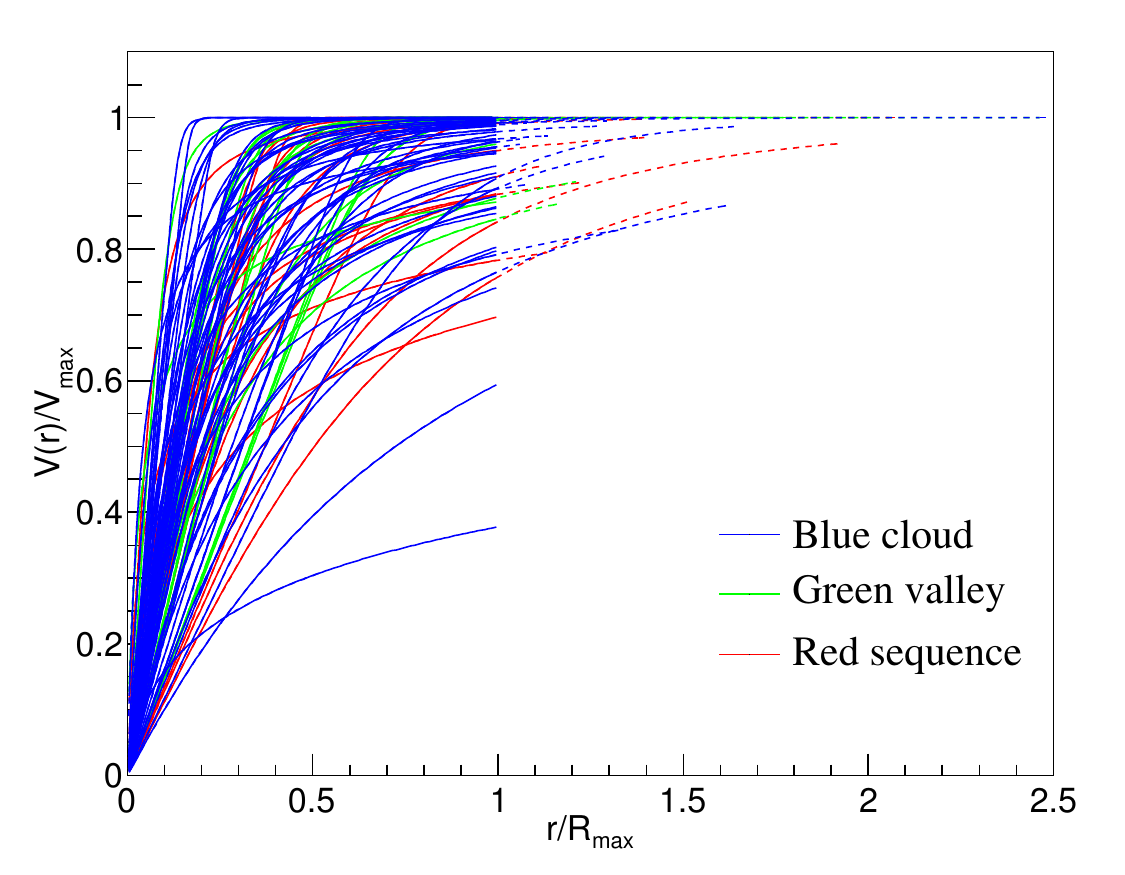}
    \includegraphics[width=0.49\textwidth]{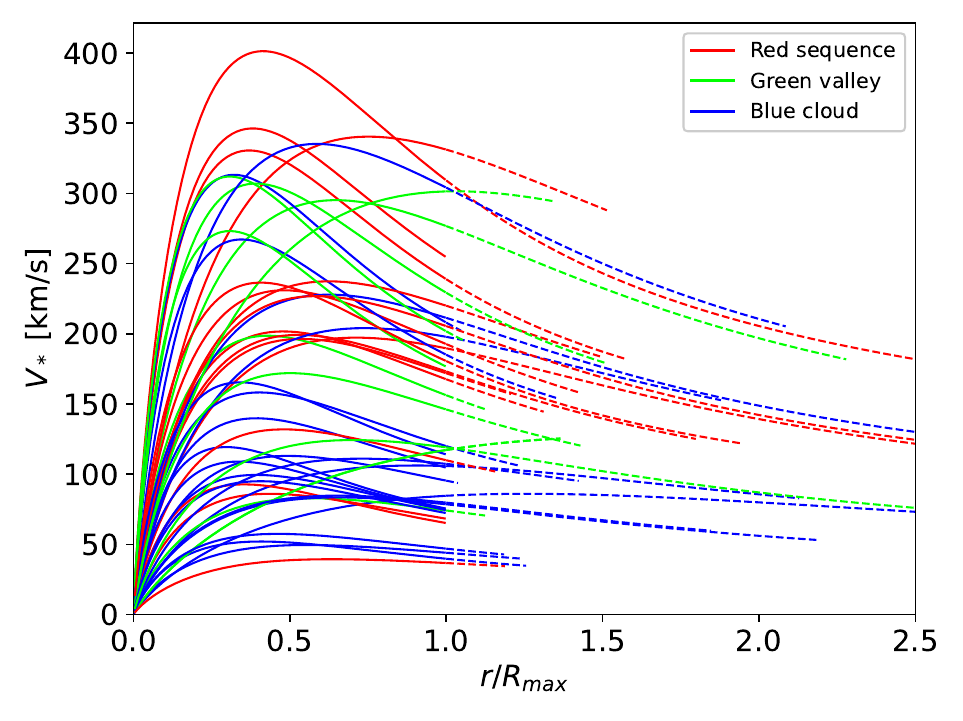}
    \caption{A sample of randomly selected rotation curves (left) and stellar 
    mass rotation curves (right) for the blue cloud (blue), green valley 
    (green), and red sequence (red) galaxies.  The solid lines extend to 
    $r = R_\text{max}$, while the dashed lines show the curves extrapolated to 
    $r = R_{90}$.}
    \label{fig:Rot_curves}
\end{figure*}

\begin{figure*}
    \centering
    \includegraphics[width=0.48\textwidth]{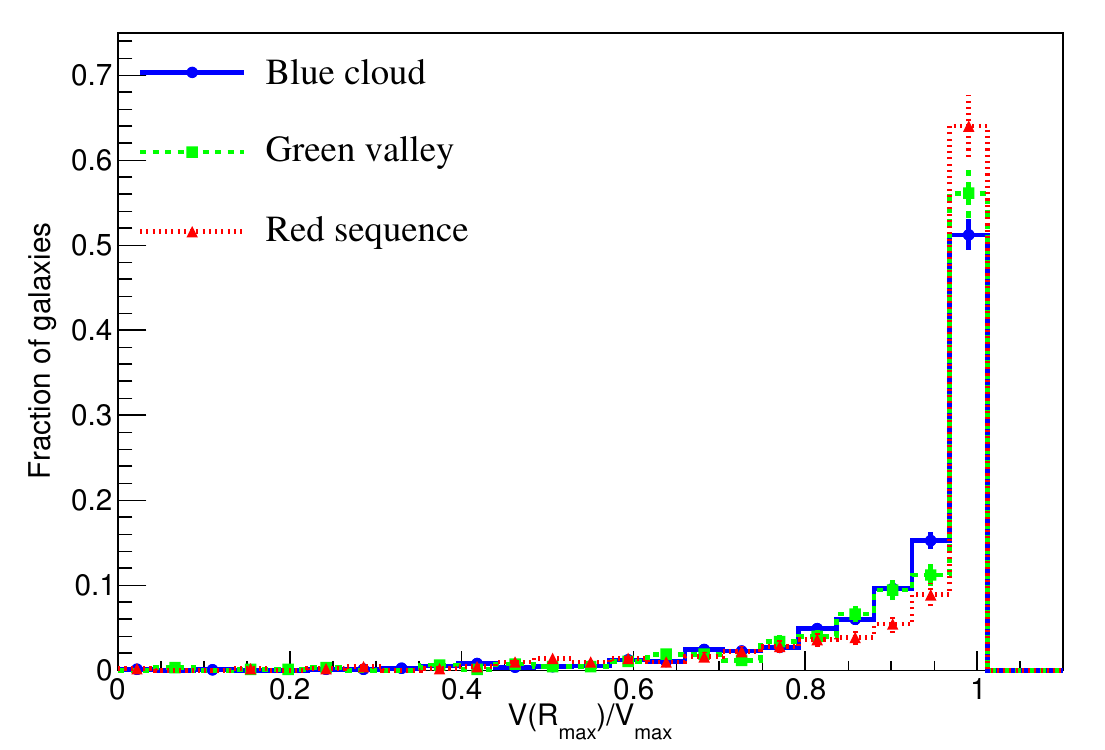}
    \includegraphics[width=0.48\textwidth]{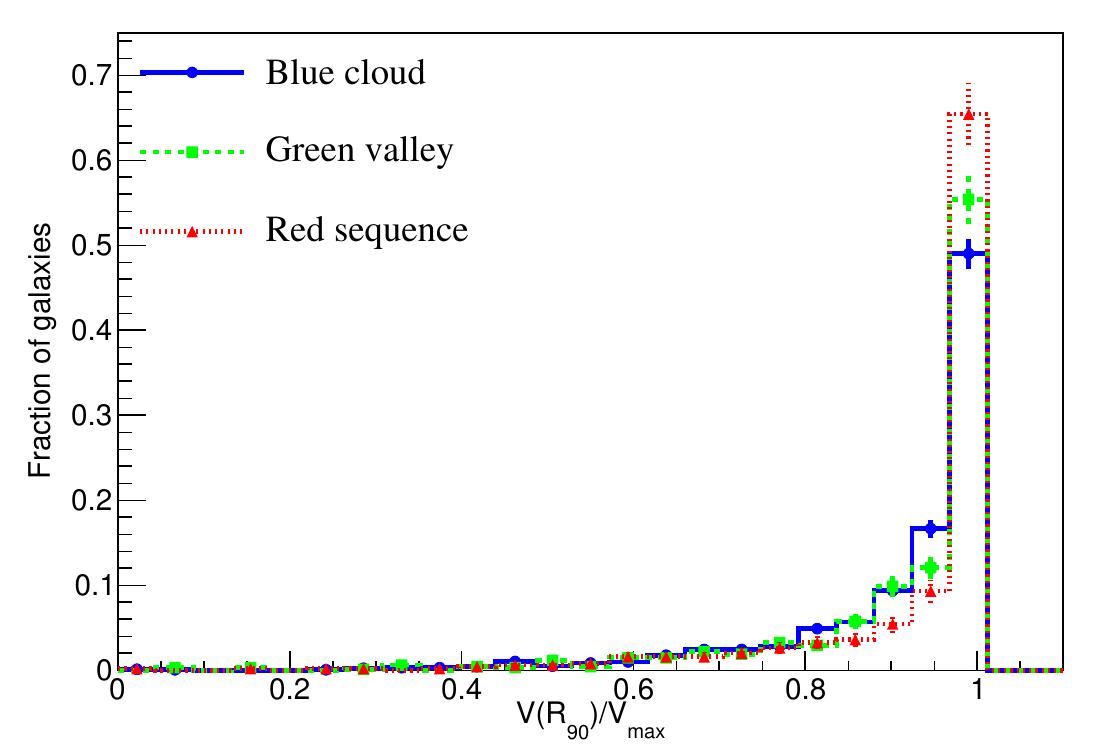}
    \caption{The distribution over the rotational velocity, $V_r (r)$, evaluated 
    at $R_\text{max}$ (left) and at $R_{90}$ (right) normalized by 
    $V_\text{max}$ for the blue cloud (blue solid line), green valley (green 
    dashed line), and red sequence (red dotted line) galaxies.}
    \label{fig:relV}
\end{figure*}

\begin{figure}
    \centering
    \includegraphics[width=0.49\textwidth]{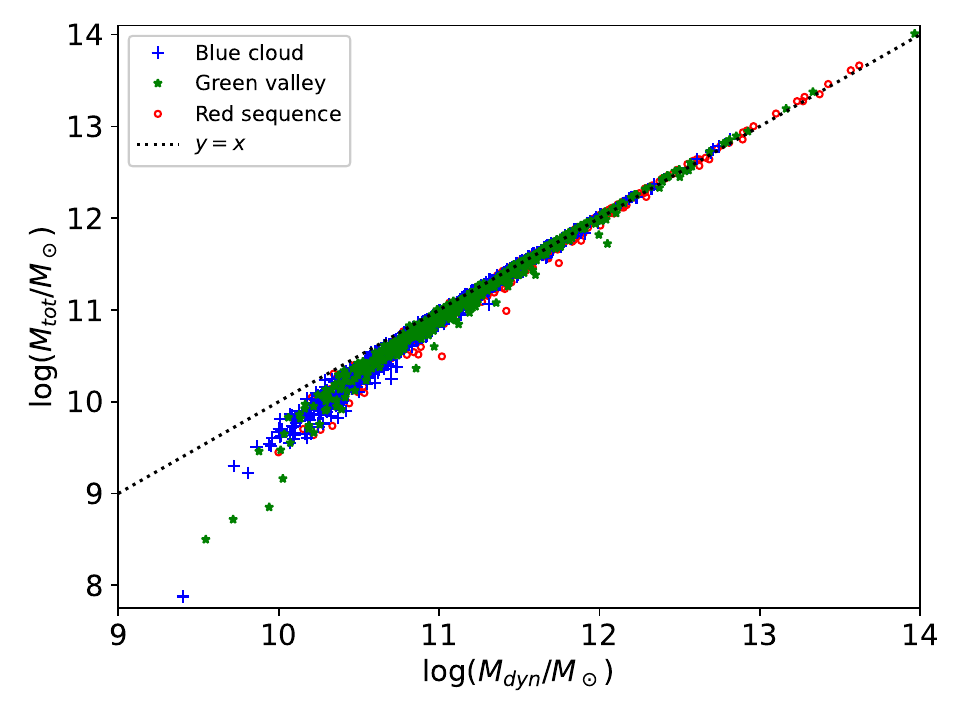}
    \caption{Relationship between the total mass, $M_\text{tot}$, and the 
    dynamical mass, $M_\text{dyn}$, calculated using the kinematic parameter 
    $S_{0.5}$, as defined in \cite{AquinoOrtiz18}.  The black dotted line 
    represents the one-to-one relationship.  We find good agreement with this 
    other mass indicator, with our $M_\text{tot}$ smaller than $M_\text{dyn}$ at 
    low mass.}
    \label{fig:M_SK_comp}
\end{figure}

We assume that a galaxy's rotational motion is dominated by Newtonian orbital 
mechanics: the orbital velocity of a particle some distance $r$ from the center 
of the galaxy is a function of the total mass internal to that radius, $M(r)$, 
assuming axial symmetry.  For spiral galaxies, the orbital motion is assumed to 
be circular.  The gravitational force is the source of the centripetal 
acceleration for a particle in orbit, so
\begin{equation}\label{eqn:M_within_r}
    M(r) = \frac{V(r)^2 r}{G}
\end{equation}
where $V(r)$ is the rotational velocity at a distance $r$ from the center of the 
galaxy and $G = 6.67408\times 10^{-11}$~m$^3$ kg$^{-1}$ s$^{-2}$ is the 
Newtonian gravitational constant.  Thus, by measuring $V(r)$ and $r$, we can 
estimate $M(r)$.  Due to the limited extent of the MaNGA H$\alpha$ velocity 
maps, rotational velocities can only be measured out to the visible extent of 
the galaxy.  As seen in Figure \ref{fig:IFU}, the IFU coverage limits the 
maximum radius, $R_\text{max}$, to which the rotational curves are evaluated.  
Figure~\ref{fig:Fr_Rmax} (left) shows the distribution over $R_\text{max}$.  We 
find that the maximum radius can be as high as 25~\hkpc, but the majority of the 
galaxies have data out to about 5~\hkpc.

To help alleviate this observational bias introduced by the size of the IFU, we 
use the parametrization of Eqn.~\ref{eqn:rot_curve} to extrapolate the rotation 
curve out to $R_{90}$, the distribution over which is shown 
Figure ~\ref{fig:Fr_Rmax} (right).  We then evaluate the galaxy's total mass, 
$M_\text{tot}$, at this radius using Eqn.~\ref{eqn:M_within_r}, as illustrated 
on the left in Figure~\ref{fig:Rot_curves} for a sample of randomly selected 
galaxies.  In Figure~\ref{fig:relV}, we present the distribution of the 
rotational velocity at $R_\text{max}$ (left) and at $R_{90}$ (right) normalized 
by the best-fit value obtained for $V_\text{max}$.  It is apparent that the 
rotational curves are closer to the plateau in the latter case.  To ensure that 
the plateau is reached in further analysis we require that 
$V(R_{90})/V_\text{max} > 0.90$.  Not only does $R_{90}$ typically extend 
further than $R_\text{max}$, as shown in Figure~\ref{fig:Fr_Rmax}, but 
evaluating the mass at $R_{90}$ allows us to consistently probe the same region 
of each galaxy.

Previous studies \citep{Cappellari06, Weiner06, AquinoOrtiz18} experimented with 
combining the rotation and random motion that support galaxies to better 
estimate the total dynamical mass of a galaxy.  Defined as
\begin{equation}
    S_K^2 = KV_\text{max}^2 + \sigma^2
\end{equation}
where $K = 0.5$, $V_\text{max}$ is the maximum rotational velocity measured in 
the galaxy, and $\sigma$ is the average velocity dispersion in the galaxy, this 
quantity is theorized to be proportional to the galaxy's total mass, 
\begin{equation}\label{eqn:Mdyn_SK}
    M_\text{tot} = \eta \frac{S_{0.5}^2r}{G}
\end{equation}
where $\eta \approx 1.8$ for $r = R_e$ \citep{AquinoOrtiz18}.  We compare our 
estimates for the total mass, $M_\text{tot}$, with the dynamical mass estimated 
with Eqn.~\ref{eqn:Mdyn_SK} evaluated at $R_{90}$.  As seen in 
Figure~\ref{fig:M_SK_comp}, we find good agreement between our results and the 
masses estimated with $S_{0.5}$.  Those galaxies with the largest difference 
between these two masses are those with the smallest masses, 
$\log(M_\text{tot}/M_\odot) \lesssim 10.5$.  These objects have very little 
velocity dispersion, indicating that they are dominated by rotation.  This 
disagreement is therefore likely due to a different value of $\eta$ or $K$ for 
these objects.  Investigating the dynamical mass estimated with $S_K$ with our 
sample is beyond the scope of this study.

\subsection{Stellar mass, $M_*$}

\begin{figure*}
    \centering
    \includegraphics[width=0.48\textwidth]{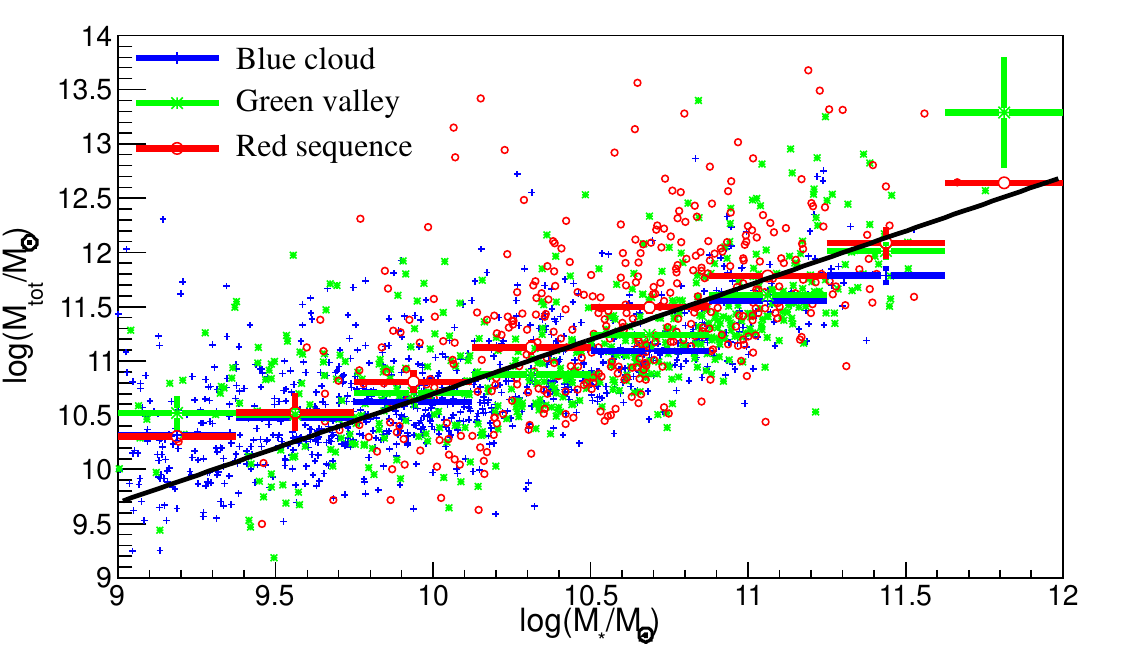}
    \includegraphics[width=0.48\textwidth]{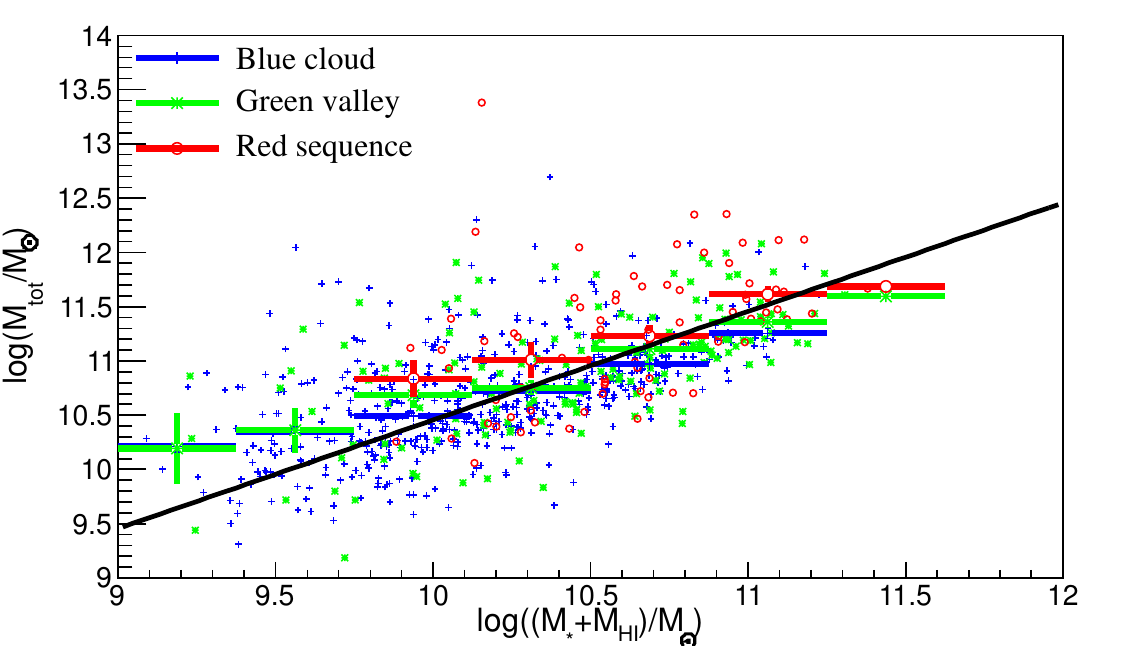}
    \caption{The relationship between the total mass and the stellar mass (left) 
    and the visible ($M_* +$ \MHI) mass (right) for each of the galaxies in our 
    sample (points).  The symbols correspond to the color-magnitude diagram 
    classification: blue crosses are for the blue cloud, green asterisks for the 
    green valley, and red open circles for the red sequence.  Galaxies in each 
    of these three populations are binned according to their stellar mass (left) 
    or visible mass (right); the median total mass in each of these bins is 
    shown in the points with error bars.  The black line is a fit to the entire 
    sample with a fixed slope of one and intercept of $0.70\pm0.01$ (left) and 
    $0.46\pm0.01$ (right).}
    \label{fig:Mtot_Mstar}
\end{figure*}

We estimate the total stellar mass, $M_*$, contained within $R_{90}$ by 
extrapolating our fitted disk rotation curve out to $R_{90}$.  We then evaluate 
the galaxy's total stellar mass, $M_*$, at this radius using 
Eqn.~\ref{eqn:M_disk}, as shown on the right in Figure~\ref{fig:Rot_curves}.  
The relationship between the stellar mass and total mass within $R_{90}$ for our 
sample of galaxies is shown on the left in Figure~\ref{fig:Mtot_Mstar}.  The 
points with error bars show the median total mass in each stellar mass bin for 
galaxies in each of the three evolutionary populations: blue cloud, green 
valley, and red sequence.

For galaxies with stellar masses above $10^{10}$ $M_\odot$, we find that the 
total mass is approximately proportional to stellar mass, similar to the results 
of \cite{Ouellette17,AquinoOrtiz20}.  We illustrate this by performing a linear 
fit to the data with a fixed slope of 1.  The result of the fit is shown by a 
solid black line with the $y$-intercept of $0.70\pm0.01$.  Galaxies in each of 
the three evolutionary populations follow this relationship, though there is 
some deviation from it for lower stellar masses.  Given the strong correlation 
between the total and stellar mass, we conclude that the ratio \MratioS is an 
appropriate variable to study.

\subsection{Neutral hydrogen, \MHI}

We investigate the effect of the \HI mass on the visible mass by combining it 
with the total stellar mass to estimate the total visible mass in a galaxy, 
$M_\text{vis}$.  Due to the limited progress of the \HI-MaNGA survey, only \NHI 
galaxies in our sample currently have an \HI detection.  The relationship 
between the total mass and the visible mass that includes \HI is presented on 
the right in Figure~\ref{fig:Mtot_Mstar}. The proportionality between these two 
quantities still holds, as is demonstrated by the linear fit with a fixed slope 
of 1 and a $y$-intercept of $0.46\pm0.01$.  As a result, we use the ratio 
$M(R_{90})/(M_*(R_{90}) +$ \MHI) = \Mratio in our study.

\subsection{Statistical model for \Mratio}\label{sec:stat_model}

\begin{figure*}
    \centering
    \includegraphics[width=\textwidth]{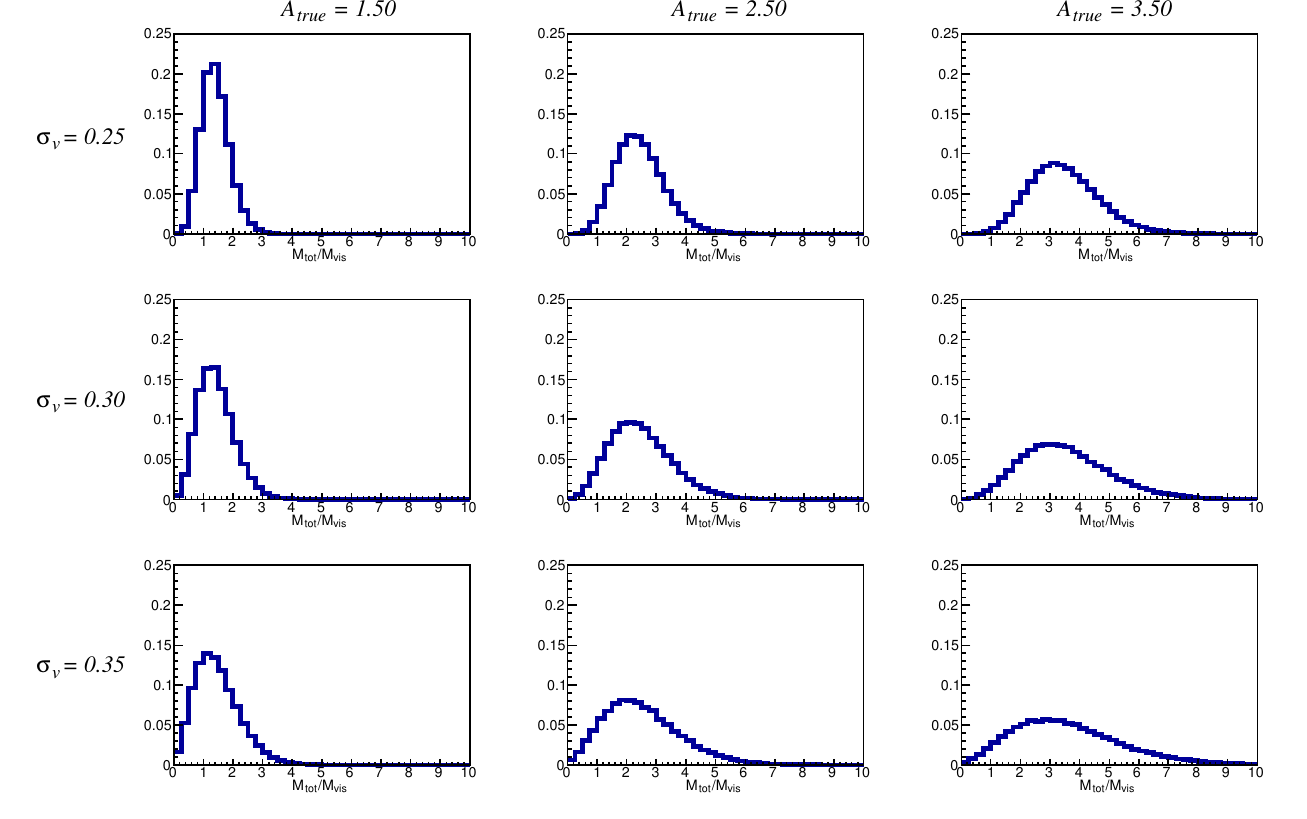}
    \caption{Templates in \Mratio with different values of $A_\text{true}$ 
    (changing horizontally from 1.0 to 2.0) and $\sigma_v$ (changing vertically 
    from 0.25 to 0.35).}
    \label{fig:Templates}
\end{figure*}

The accuracy of determining \Mratio is limited by the statistical and systematic 
uncertainties on the rotational velocity at the 90\% elliptical Petrosian 
radius, $V(R_{90})$.  We develop a statistical model that accounts for the 
resolution on the rotational velocity and uncovers the true value of \Mratio, 
referred to as $A_\text{true}$.  This model uses the data distribution in the 
observed \Mratio to fit for the best value of the relative uncertainty on the 
rotational velocity $\sigma_v$ together with $A_\text{true}$, which should be 
understood as a mean true value for a sample of galaxies included in the fit.

We assume that the best-fit maximum velocity $V_\text{max, obs}$ is normally 
distributed around its true value $V_\text{true}$ with the standard deviation of 
$\sigma_v V_\text{true}$.  We also assume that $M_\text{tot}$ is equal to the 
product of the measured value of the visible mass, $M_\text{vis}$, and 
$A_\text{true}$.  We construct template distributions in \Mratio for several 
values of $A_\text{true}$ and $\sigma_v$.  To build a given template, we 
multiply each galaxy's $M_\text{vis}$ by the template's value of $A_\text{true}$ 
to estimate the galaxy's total mass.  From the total mass, we calculate the true 
value of the velocity, $V_\text{true}$, using the relation given in 
Eqn.~\ref{eqn:M_within_r} with $r = R_{90}$.  $V_\text{true}$ is then smeared 
according to a normal distribution with standard deviation 
$\sigma = V_\text{true}\sigma_v$, where $\sigma_v$ is the value of the relative 
uncertainty on the rotational velocity assumed in this template.  This normal 
smearing is repeated 100 times to construct smoother templates.  The smeared 
velocity, $V_\text{obs}$, is then converted back into an ``observed'' total 
mass, from which we then calculate the ``observed'' mass ratio \Mratio.  The 
procedure is repeated for all galaxies in a given sample. Thus, each template 
histogram has 100 times more entries than the data histogram.

For illustration, we show an example 3$\times$3 template matrix in 
Figure~\ref{fig:Templates}.  Since $M_\text{tot}$ depends on $V_\text{max}^2$, a 
symmetric distribution in $V_\text{max}$ results in an asymmetric distribution 
in the observed \Mratio with a preference for larger values of \Mratio.  As can 
be seen from these plots, larger values of $A_\text{true}$ result in the peak of 
the distribution being shifted towards larger values of the observed \Mratio, 
while larger values of $\sigma_v$ result in a broadening of the distributions.  
Thus, comparing the templates with the data distributions constrains both 
parameters. 

We construct templates for seven different values of $A_\text{true}$ and seven 
different values of $\sigma_v$, resulting in 49 templates.  To identify which 
combination of $A_\text{true}$ and $\sigma_v$ best represents the data, we 
evaluate the agreement of each template with the data distribution using a 
$\chi^2_\nu$ criterion.  $\chi^2_\nu$ is normalized by the number of the degrees 
of freedom, which is equal to the number of bins minus the three free parameters 
from the fit (normalization, $A_\text{true}$ and $\sigma_v$).  We keep 
$\sigma_v$ as a free parameter of the fit to account for variation from sample 
to sample.  The observed optimal values of $\sigma_v$ vary from 0.17 to 0.3.

The advantage of this statistical model is that it constrains the uncertainties 
on the rotational velocity \emph{in situ}.  Since the uncertainty on the 
rotational velocity affects the width of the distribution in the mass ratio, the 
uncertainty on the velocity (either statistical or systematic) is constrained by 
the fit itself.

Previous observations of the stellar-halo mass relation (SHMR) show that \Mratio 
varies among a galaxy population \citep[and references therein]{Wechsler18}, so 
the fitted value of $A_\text{true}$ should be understood as an average value of 
\Mratio in a given population. The developed statistical model provides a good 
description of the data.  In particular, a normal distribution in $V_\text{obs}$ 
rather than in \Mratio explains the asymmetric shape of the distribution over 
\Mratio.  Deviations from the model for large values of \Mratio can be 
attributed either to non-gaussian tails in the experimental uncertainties or to 
large deviations from the average in \Mratio in a particular population.

\section{Correlation of the mass ratio with evolutionary and environmental properties}\label{sec:correlations}

\subsection{Dependence of the mass ratio on the color-magnitude classification}\label{sec:Mratio_CMD}

\begin{figure*}
    \includegraphics[width=\textwidth]{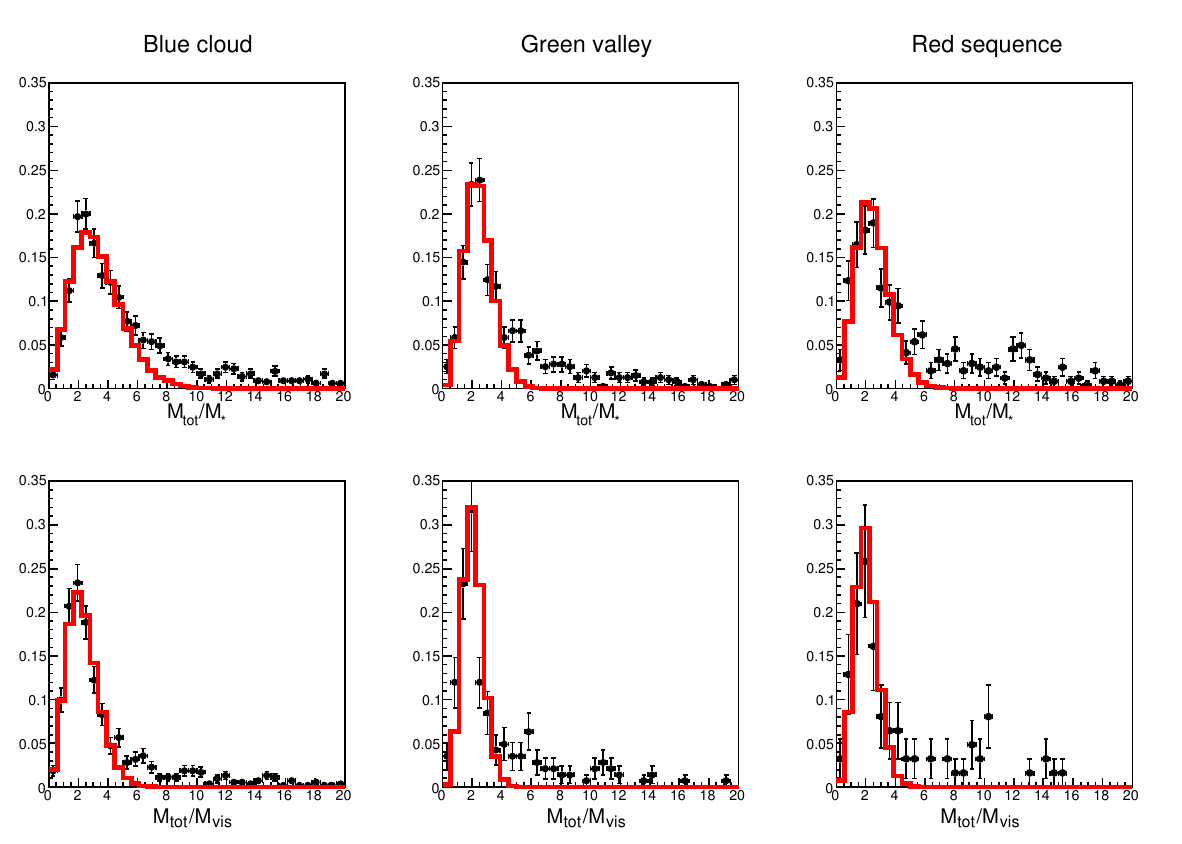}
    \caption{The distributions in \MratioS (top row) and  \MratioHI(bottom row) 
    for the sample of galaxies (points) compared to the model prediction based 
    on the best fit template (histogram) for the three galaxy populations: blue 
    cloud (left column), green valley (middle column), and red sequence (right 
    column).}
    \label{fig:MtMv_bybin_CMD}
\end{figure*}

Due to the observed differences in the relationships between $M_*$ and 
$M_\text{tot}$ for the blue cloud, green valley, and red sequence galaxies seen 
in Figure \ref{fig:Mtot_Mstar}, we perform the template fitting to these three 
populations separately.  The results of the template fit to \MratioS and \Mratio 
are shown in Figure~\ref{fig:MtMv_bybin_CMD}.  Each of these plots demonstrates 
that the statistical model provides a good description of the data 
distributions.  The best fit results and medians of the corresponding 
distributions in the observed \MratioS and \Mratio are summarized in 
Table~\ref{tab:summary}.  Medians, which are extracted from the data 
distributions without the use of the statistical model, would be equal to the 
best-fit values of $A_\text{true}$ should the data follow the model's prediction 
exactly.  The observed medians follow the same trend as the values of 
$A_\text{true}$ but are generally somewhat higher due to deviations in the data 
distributions from the model's predictions, predominantly at the high values of 
\MratioS and \Mratio (the ``tails'' of the distributions).  These deviations are 
reduced once the \HI mass is included.  Both the median and $A_\text{true}$ 
values suggest that galaxies in the blue cloud have the highest values of both 
\MratioS and \Mratio.

\begin{deluxetable*}{lcCCcCC}
    \tablewidth{0pt}
    \tablecolumns{9}
    \tablecaption{Sample statistics and fitting results for galaxy classes \label{tab:summary}}
    \tablehead{ & & \multicolumn{2}{c}{\text{\MratioS}} & & \multicolumn{2}{c}{\text{\MratioHI}} \\
    Class & Count & A_\text{true} & \text{Median} & Count with \HI & A_\text{true} & \text{Median} }
    \startdata
        Blue cloud   & 931 & 3.10\pm 0.26 & 3.91\pm 0.12 & 531 & 2.20\pm 0.20 & 2.58\pm 0.14  \\
        Green valley & 493  & 2.37\pm 0.22 & 3.02\pm 0.16 & 151  & 2.01\pm 0.37 & 2.19\pm 0.24  \\
        Red sequence & 390  & 2.4\pm 0.27 & 3.52\pm 0.23 & 77  & 2.09\pm 0.54  & 2.47\pm 0.41 \\
    \enddata
    \tablecomments{The mass ratio values here correspond to the best-fitting 
    template's $A_\text{true}$.  The median values correspond to the median of 
    the distribution of \MratioS and \Mratio in the data.}
\end{deluxetable*}

Galaxies in the blue cloud, green valley, and red sequence have different 
properties.  As shown in \cite{Salim14a,Douglass_thesis,Coenda18,Jian20}, 
galaxies in the red sequence are the brightest, most massive, and have the 
lowest star formation rates.  Blue cloud galaxies are the opposite: they are the 
faintest, the least massive, and have the highest star formation rates.  
Galaxies in the green valley tend to have luminosities and stellar masses 
comparable to those in the red sequence, but their star formation rates are 
intermediate.

Therefore, galaxies in the blue cloud are actively forming stars, while galaxies 
in the red sequence have undergone processes which have quenched their star 
formation.  Green valley galaxies are either currently undergoing these 
quenching processes (moving from the blue cloud to the red sequence), or their 
star formation is restarting (transitioning from the red sequence to the blue 
cloud).  Similar to \cite{TorresFlores11}, we observe that the blue cloud 
galaxies have the highest values of \MratioS and \Mratio.  This trend is reduced 
once \MHI is included.  These observations suggest a correlation between 
\MratioS or \Mratio and the current state of star formation.  We test this 
hypothesis by investigating the relationships between both \MratioS and \Mratio 
and several galaxy properties related to the galaxy's evolution history and its 
local environment.

\subsection{Dependence of the mass ratio on luminosity and gas-phase metallicity}\label{sec:Mr_metal}

\begin{figure*}
    \centering
    \includegraphics[width=0.49\textwidth]{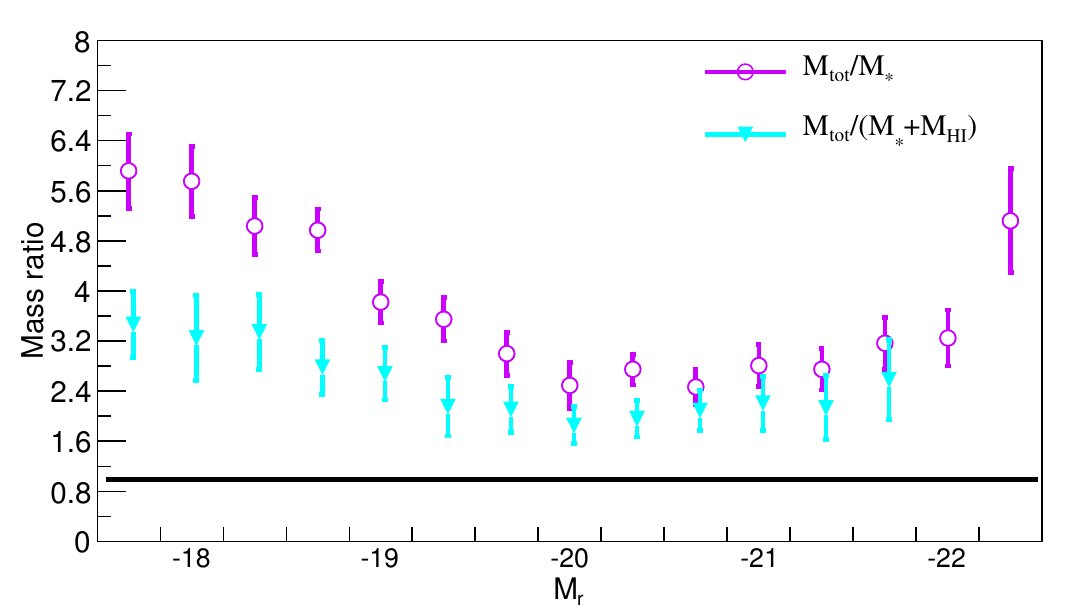}
    \includegraphics[width=0.49\textwidth]{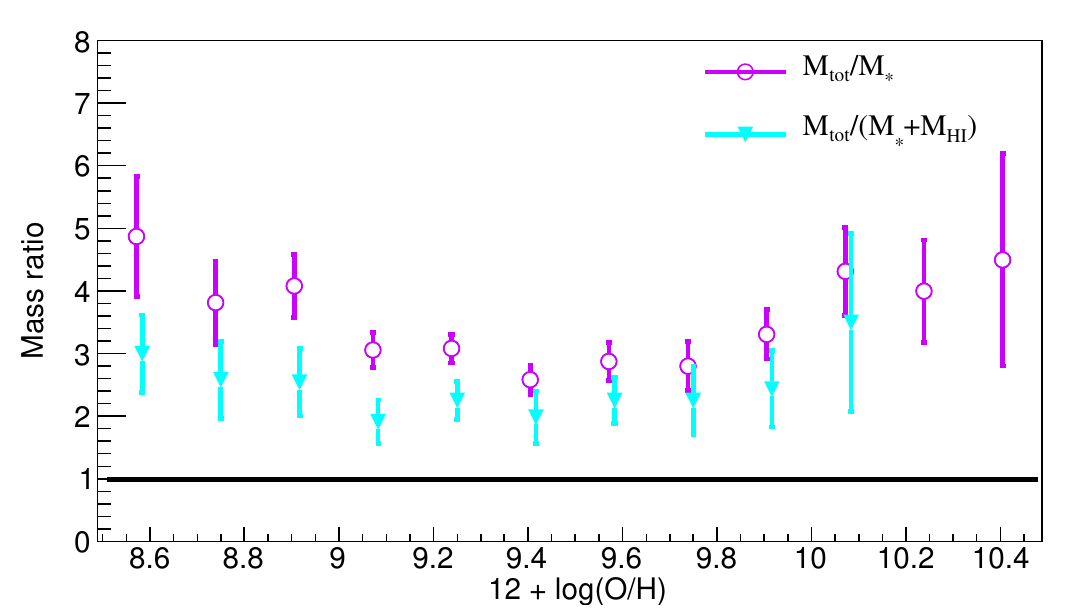}
    \caption{The dependence of median values of \MratioS (open magenta circles) 
    and \MratioHI (filled cyan triangles) on the absolute magnitude, $M_r$ 
    (left), and gas-phase metallicity, \logOH (right).  The solid black line 
    represents a ratio of 1.}
    \label{fig:MtMv_Mr_Metal_All}
\end{figure*}

\begin{figure*}
    \centering
    \includegraphics[width=0.48\textwidth]{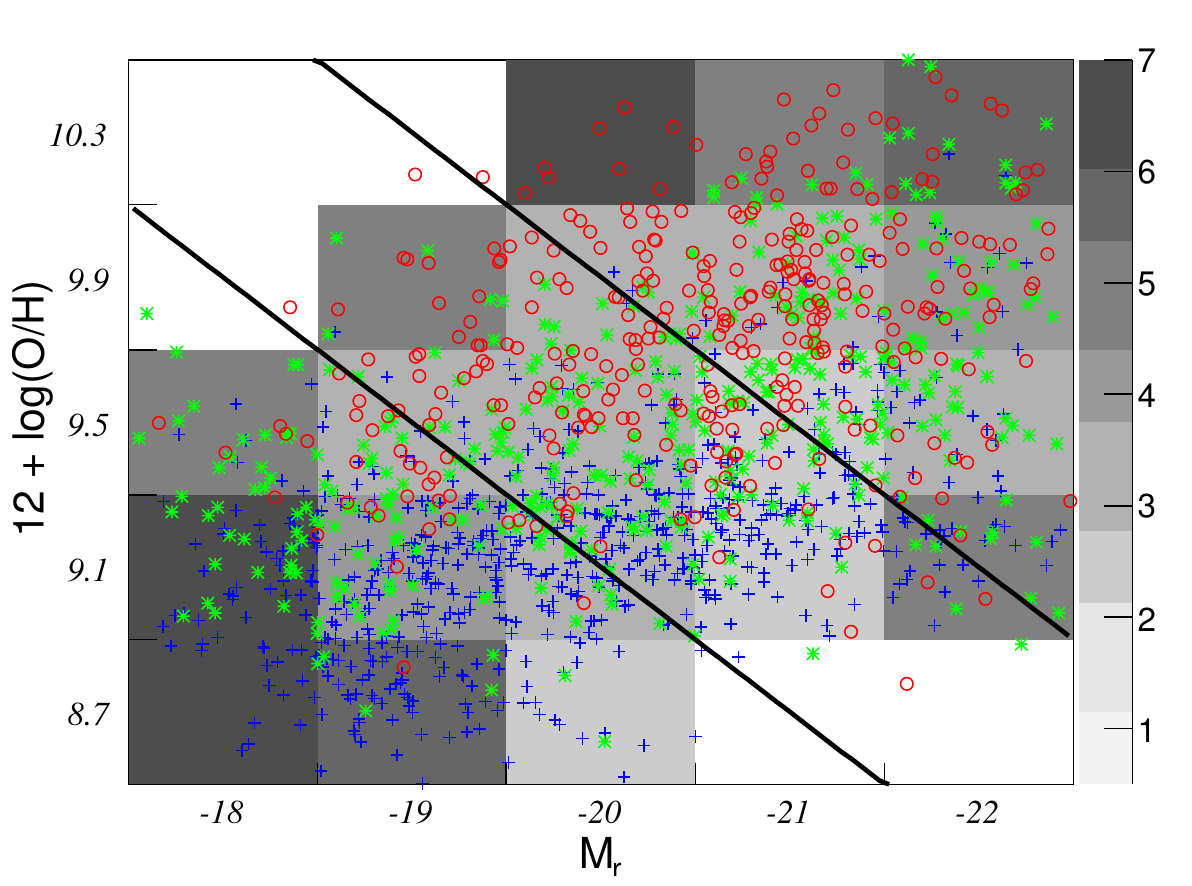}
    \includegraphics[width=0.48\textwidth]{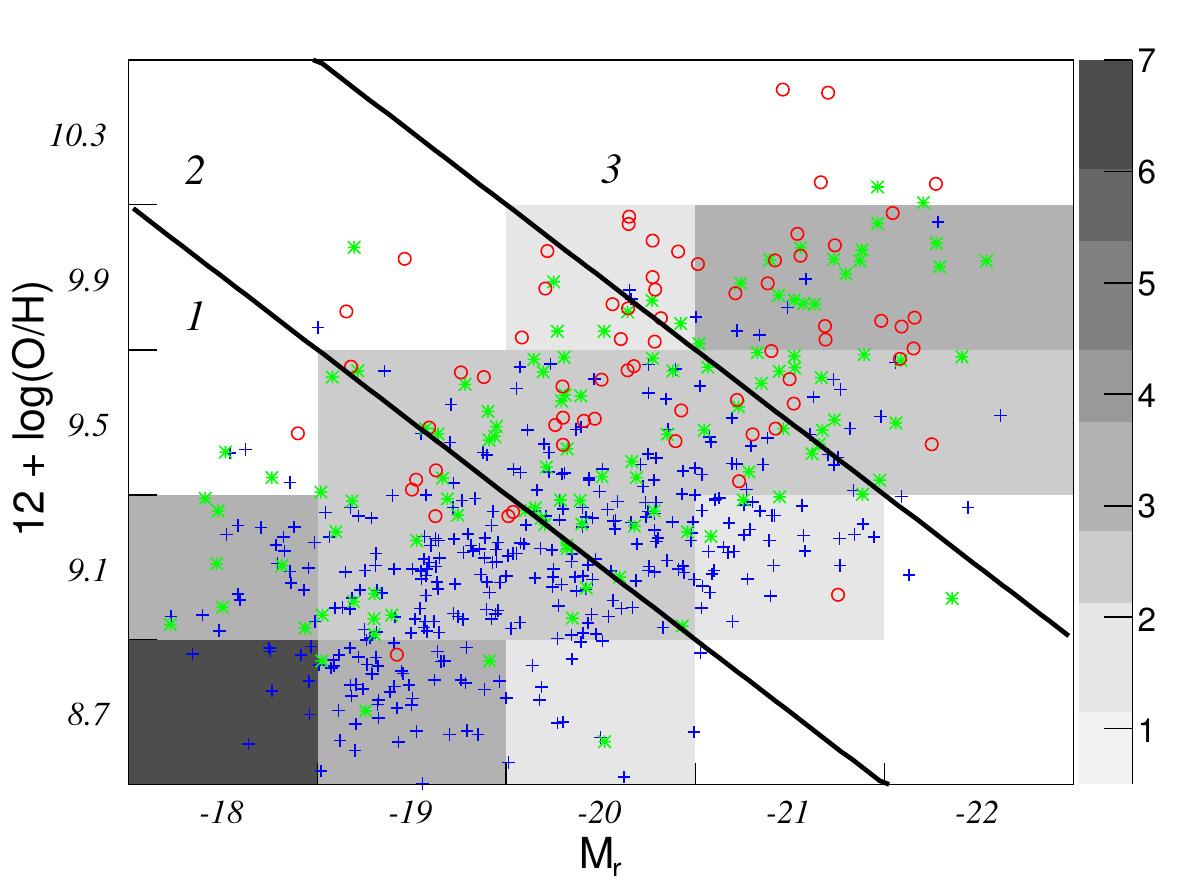}
    \caption{The grey scale shows the median values of \MratioS (left) and 
    \MratioHI (right) in the bins of $M_r$ and \logOH.  Overlaid are the 
    galaxies in the blue cloud (blue crosses), green valley (green asterisks), 
    and red sequence (red open circles).  The solid diagonal lines (\logOH 
    $= 0.4M_r + 17.1$ and \logOH $= 0.4M_r + 17.9$) indicate the boundaries for 
    the three categories, labeled by the numbers in the right panel.}
    \label{fig:MtM_Mr_Metal}
\end{figure*}

\begin{figure*}
    \centering
    \includegraphics[width=\textwidth]{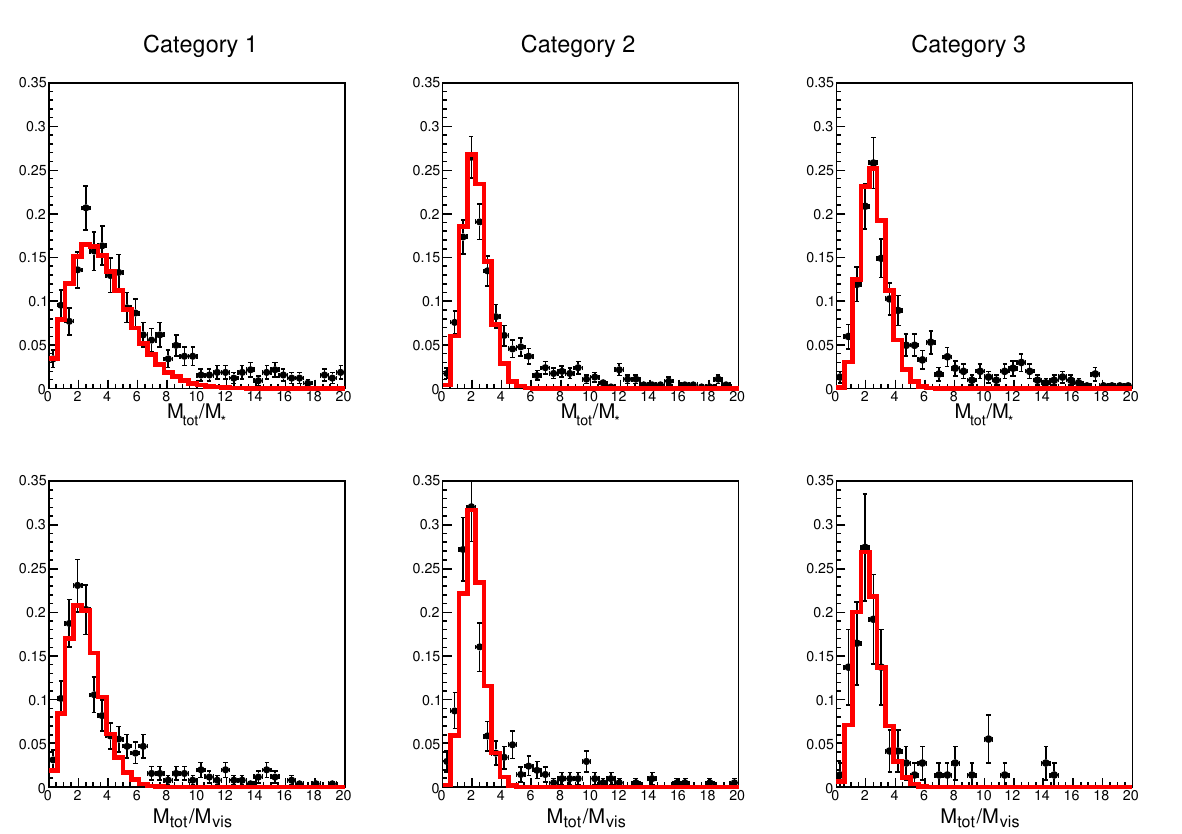}
    \caption{The distributions in \MratioS (top row) and \MratioHI (bottom row) 
    for the sample of galaxies (points) compared to the model prediction based 
    on the best fit template (histogram) for the three galaxy categories defined 
    in Figure~\ref{fig:MtM_Mr_Metal}: 1 (left column), 2 (middle column), and 3 
    (right column).}
    \label{fig:MtMs_bybin_3cats}
\end{figure*}

We study the dependence of \MratioS and \Mratio on a galaxy's luminosity, $M_r$, 
and gas-phase metallicity, \logOH, defined as the relative abundance of oxygen 
to hydrogen.  Galaxies typically follow the mass-metallicity relation 
\citep[e.g.,][]{Tremonti04}, where more massive galaxies have higher 
metallicities.  While a galaxy's metallicity should depend on its stellar mass, 
it should also increase with its total mass due to the corresponding deeper 
potential well.  As shown in \cite{Douglass_thesis}, blue cloud galaxies have 
slightly lower metallicities than those in the red sequence, indicative of being 
in an earlier stage of their star formation histories.  Similarly, red sequence 
galaxies are advanced in their star formation process and thus have higher 
metallicities.

While required to study these relationships, a fairly fine binning in $M_r$ and 
\logOH does not permit a fit to the statistical model due to insufficient 
statistics in each bin.  We instead extract the medians of the distribution in 
each bin.  The statistical model is then applied to the coarsely binned data 
based on both $M_r$ and \logOH.  On the left in 
Figure~\ref{fig:MtMv_Mr_Metal_All}, we show the dependence of the median 
\MratioS and \Mratio values on the luminosity.  We observe that the 
faintest galaxies (those with $M_r > -20$) have the largest value of mass ratio.  
This behavior agrees with prior observations by 
\cite{Persic96,Strigari08,TorresFlores11,Karukes17,Behroozi19,DiPaolo19,Douglass19} 
and the simulation results of \cite{Moster10}, who predict that faint galaxies 
are more enriched in dark matter.  Similar to these previous results, we also 
observe a rise in each of the mass ratios for the brightest galaxies, typically 
members of the red sequence type.

The relationship between each of the mass ratios and the gas-phase metallicity 
of the galaxies is shown on the right in Figure~\ref{fig:MtMv_Mr_Metal_All}.  
Similar to the correlation between both \MratioS and \Mratio and $M_r$, we 
observe a rise in both \MratioS and \Mratio for galaxies with both high and low 
metallicities.

The addition of \HI to the visible mass has a significant effect on faint 
galaxies and those with the lowest metallicities, while a smaller effect is 
observed for brighter, higher metallicity galaxies.  This is similar to the 
observations made by \cite{TorresFlores11}, who found that including \MHI in the 
total visible mass estimate reduced the deviation from the baryonic Tully-Fisher 
relation by the low-mass galaxies.

\begin{deluxetable*}{ccCCcCC}
    \tablewidth{0pt}
    \tablecolumns{9}
    \tablecaption{Sample statistics and fitting results for galaxy categories
    \label{tab:3cats}}
    \tablehead{ & & \multicolumn{2}{c}{\MratioS} & & \multicolumn{2}{c}{\MratioHI} \\
    Category & Count & A_\text{true} & Median & Count with \HI & A_\text{true} & Median }
    \startdata
        1  & 495 & 3.36\pm 0.46 & 4.25\pm 0.17 & 255 & 2.35\pm 0.31 & 2.64\pm 0.20 \\
        2  & 517 & 2.20\pm 0.21 & 2.67\pm 0.14 & 207 & 2.05\pm 0.27 & 2.07\pm 0.20 \\
        3  & 427  & 2.50\pm 0.26 & 3.15\pm 0.19 & 81  & 2.15\pm 0.46 & 2.32\pm 0.33 \\
    \enddata
    \tablecomments{Categories are as defined in Figure \ref{fig:MtM_Mr_Metal}.  
    All galaxies included in these fits are required to have gas-phase 
    metallicity estimates.  The mass ratio values here correspond to the 
    best-fitting template's $A_\text{true}$.  The median values correspond to 
    the median of the distribution of \MratioS and \Mratio in the data.}
\end{deluxetable*}

We further examine the dependence of both \MratioS and \Mratio on the luminosity 
and gas-phase metallicity by plotting each of the mass ratios in the 2D plane of 
($M_r$, \logOH), shown in Figure~\ref{fig:MtM_Mr_Metal}.  We find that the 
highest values of \Mratio are observed in faint galaxies with low metallicities 
and bright galaxies with high metallicities.  To isolate these trends, we divide 
the galaxies into three categories as marked in the right panel of 
Figure~\ref{fig:MtM_Mr_Metal}, and perform a fit to the statistical model 
described in Section~\ref{sec:stat_model} to find the average value of both 
\MratioS and \Mratio in each of these three categories.  The results of this 
fitting are presented in Figure~\ref{fig:MtMs_bybin_3cats} and summarized in 
Table~\ref{tab:3cats}.

It is apparent that higher values of both \MratioS and \Mratio are observed for 
faint galaxies with low metallicities (category 1), which is mostly populated by 
galaxies in the blue cloud.  There is also an increase in both \MratioS and 
\Mratio for the brightest galaxies with the highest metallicities (category 3), 
populated predominantly by galaxies in the red sequence and green valley.  
Galaxies populating the central region of Figure~\ref{fig:MtM_Mr_Metal} 
(category 2) have the lowest values of both \MratioS and \Mratio.  When 
comparing the top and bottom rows of Figure~\ref{fig:MtMs_bybin_3cats}, we see 
that the addition of \HI significantly reduces the tails in the distribution 
over the mass ratio, especially for those galaxies in category 1.  This observed 
dependence on $M_r$ and \logOH suggests a correlation of \Mratio with a galaxy's 
star formation history.

\subsection{Estimating \MratioS based on redshift and photometry}
\label{sec:eval}

\begin{figure*}
    \centering
    \includegraphics[width=0.49\textwidth]{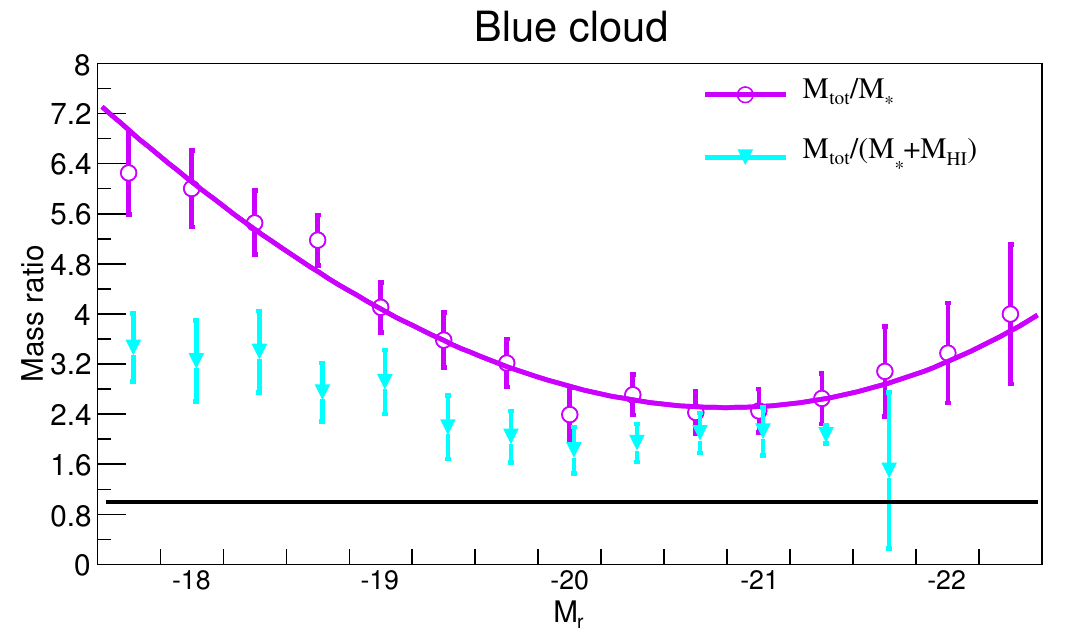}
    \includegraphics[width=0.49\textwidth]{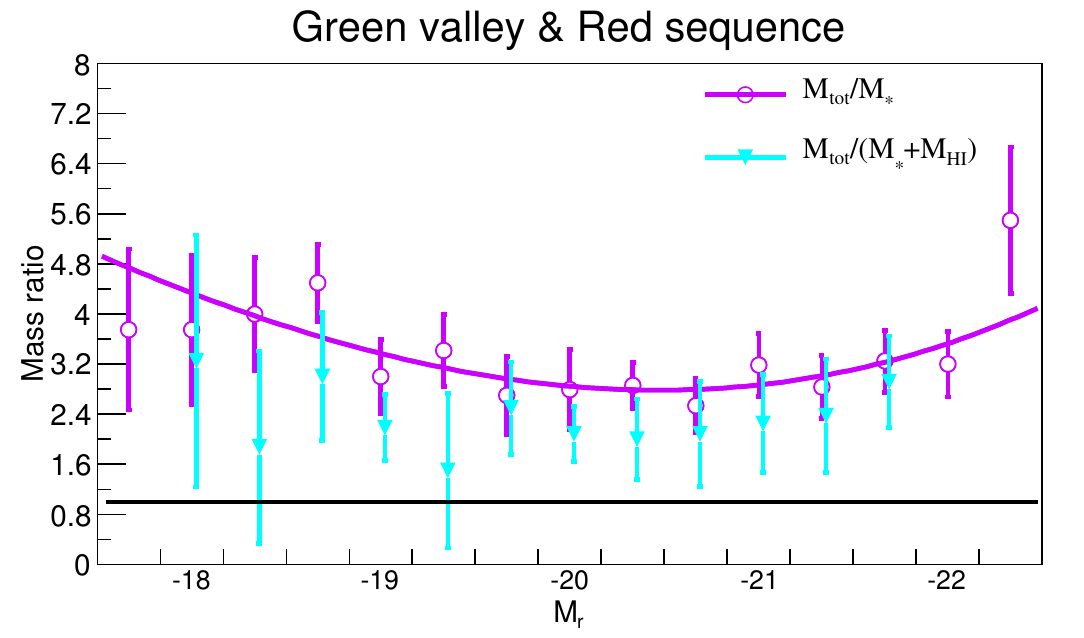}
    \caption{The dependence of median values of \MratioS (open magenta circles) 
    and \MratioHI (filled cyan triangles) on $M_r$ for galaxies in the blue 
    cloud (left panel) and in the green valley and red sequence (right panel).  
    The solid magenta lines show the best fits of \MratioS to a third power 
    polynomial.}
    \label{fig:MtMv_Mr_fit}
\end{figure*}

\begin{figure}
    \includegraphics[width=0.5\textwidth]{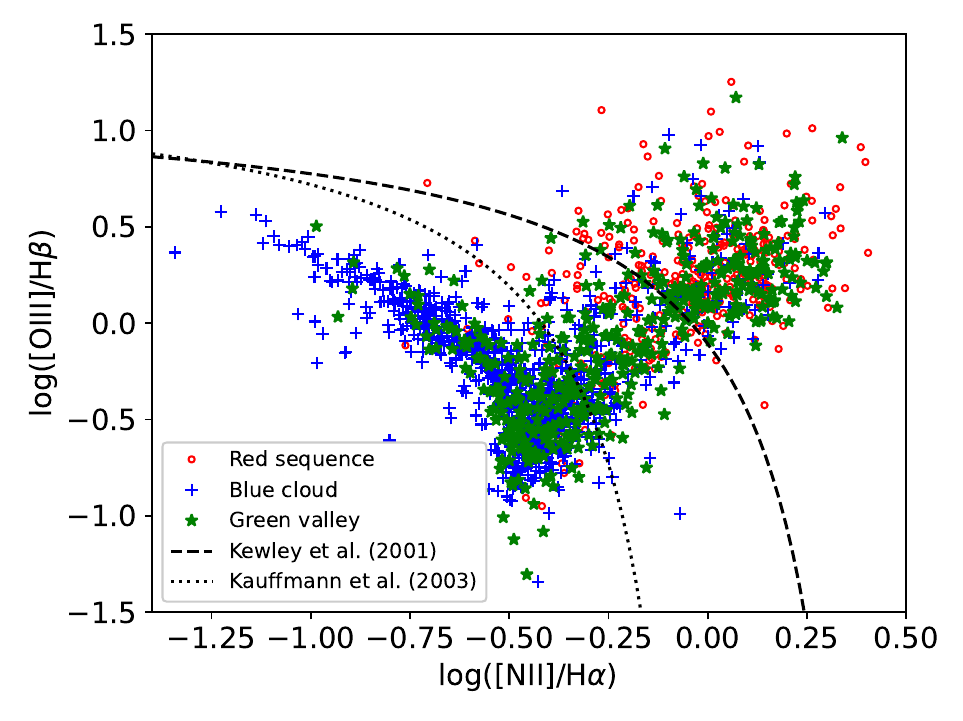}
    \caption{A BPT diagram for our sample of SDSS MaNGA galaxies, marked by 
    their color-magnitude diagram classification: open red circles for the red 
    sequence, green stars for the green valley, and blue crosses for the blue 
    cloud.  Objects below the dotted line defined by \cite{Kauffmann03} are 
    considered to have their main source of ionization from star formation, 
    while objects above the dashed line defined by \cite{Kewley01} are expected 
    to have their main source of ionization from AGN activity.  Galaxies in 
    between the two lines are composite galaxies, where their ionization comes 
    from both star formation and AGN activity.}
    \label{fig:BPT}
\end{figure}

There is a class of problems, e.g. Large Scale Structure (LSS) studies, where it 
is important to evaluate the ratio of halo to stellar mass based on some easily 
accessible observables.  Due to the systematics and relatively high S/N spectra 
required for its calculation, the gas-phase metallicity is not easily 
obtainable.  However, a galaxy's luminosity and its color-magnitude 
classification can be easily evaluated based on photometry and a redshift.  
Thus, we derive a parametrization of \MratioS based on these two observables.  
Because of the limited statistics in our current sample, we combine the green 
valley and red sequence galaxies for this parameterization.  Fits to the third 
power polynomials for these two populations of galaxies (blue cloud and red 
sequence / green valley) are shown in Figure~\ref{fig:MtMv_Mr_fit}.  The results 
of the best fits are summarized in Table~\ref{tab:Mr_fit}.

\begin{deluxetable*}{cCCCC}
    \tablewidth{0pt}
    \tablecolumns{5}
    \tablecaption{\label{tab:Mr_fit}Parameterization of \MratioS}
    \tablehead{ Class & p_0 & p_1 & p_2 & p_3 }
    \startdata
     Blue cloud   & 33.8\pm 1.1 & 6.10\pm 0.07 & -0.80\pm 0.004 & 0.021\pm 0.0001\\
     Green valley \& Red sequence & -15.2\pm 1.5 & 8.66\pm 0.09 & -0.72\pm 0.004 & 0.017\pm 0.0002\\
    \enddata
    \tablecomments{Values of the best fit parameters to 
    \MratioS $= p_0 + p_1 x + p_2 x^2 + p_3 x^3$, where $x = M_r$.}
\end{deluxetable*}

This version of the stellar-halo mass relation (SHMR) exhibits its key 
characteristics: the fraction of dark matter decreases with increasing stellar 
mass for all but the brightest galaxies 
\citep[and others]{Wechsler18,Douglass19}.  As expected, galaxies in the blue 
cloud show a strong rise in \MratioS at the faint end, while galaxies in the 
green valley and red sequence exhibit an increase in \MratioS for both the 
brightest and faintest galaxies.  Contrary to \cite{Behroozi19}, we find that 
the brightest quenched/quenching galaxies (galaxies in the green valley and red 
sequence) exhibit similar \MratioS to their star-forming counterparts.  We also 
find that faint galaxies in the blue cloud have significantly more dark matter 
compared to galaxies of similar luminosity in the green valley and red sequence, 
consistent with the results of \cite{AquinoOrtiz20}.  The addition of \MHI has 
the most significant impact on these galaxies.

Various feedback processes that reduce the star formation efficiency are often 
cited as the source for the divergence from a constant SHMR: supernovae feedback 
in fainter galaxies, and AGN feedback in the brightest galaxies 
\citep{Wechsler18,DiPaolo19}.  With galaxies in the blue cloud having higher 
star formation rates, we would expect these to behave more similar to the faint 
galaxies, and we can hypothesize that their deviation from a constant SHMR is 
largely due to supernovae feedback.  On the other hand, galaxies in the green 
valley and red sequence have much lower star formation rates; their feedback is 
likely due to AGN, which is also potentially responsible for their quenched star 
formation.

To help probe the source of feedback in these galaxies, we look at where the 
galaxies fall in the Baldwin-Phillips-Terlevich \citep[BPT;][]{Baldwin81} 
diagram.  The relative amount of [O{\sc iii}] $\lambda$ 5007 / H$\beta$ to 
[N{\sc ii}] $\lambda$ 6854 / H$\alpha$ determines whether a galaxy's ionization 
source is primarily from star formation, AGN activity, or a combination of the 
two.  We define the boundary between star-forming and composite galaxies as 
given in \cite{Kauffmann03}, and the boundary between composite and AGN galaxies 
is as defined by \cite{Kewley01}.  The BPT diagram for our sample of galaxies 
is shown in Figure~\ref{fig:BPT}, where we see that most of the galaxies 
classified as being in the blue cloud are star-forming, while the red sequence 
objects are predominantly composite galaxies and AGN.  Galaxies in the green 
valley span all three populations.  This supports the hypothesis that the 
fainter, more metal-poor galaxies in the blue cloud (located on the upper left 
of the BPT diagram) deviate from a constant SHMR because of supernova feedback, 
while the feedback in the most massive, metal-rich galaxies in the red sequence 
and green valley (the upper right of the BPT diagram) is due to AGN activity.

\subsection{Dependence of the mass ratio on the distance to the nearest neighbor}\label{sec:DNN}

\begin{figure}
    \centering
    \includegraphics[width=0.5\textwidth]{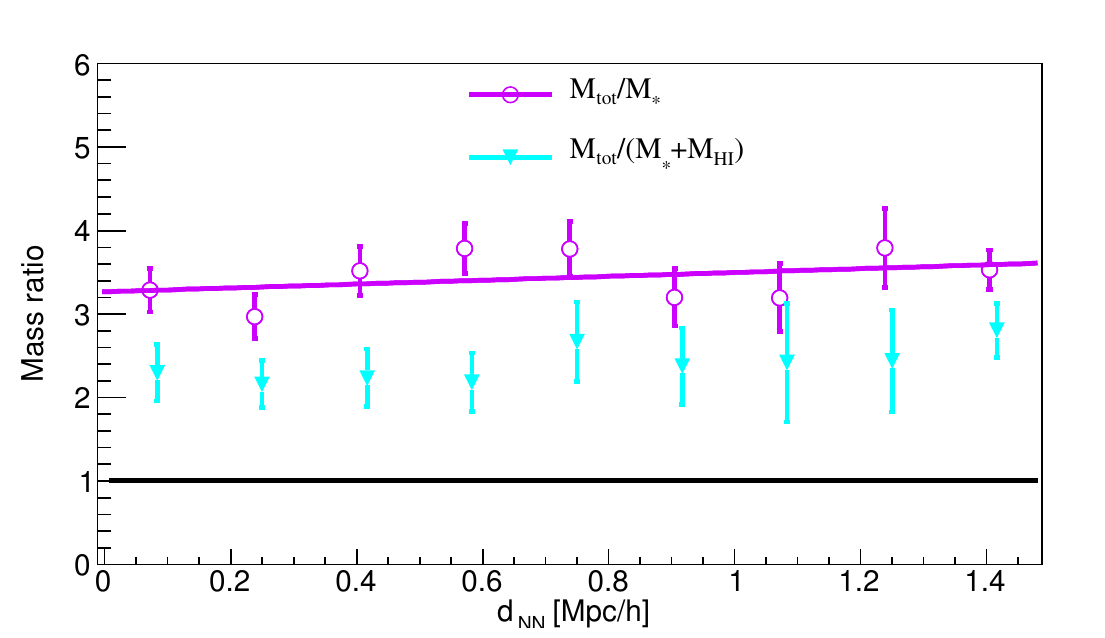}    
    \caption{The relationship between the median values of \MratioS (open 
    magenta circles) and \MratioHI (filled cyan triangles) and the distance to 
    the nearest neighbor.  The linear fit to \MratioS is shown in the solid 
    magenta line, with a slope of $0.23\pm 0.22$ and a $y$-intercept of 
    $3.27\pm 0.2$.}
    \label{fig:MtMv_DNN_All}
\end{figure}

We investigate the effect of the local environment on both \MratioS and \Mratio 
by looking at the relationship between a galaxy's mass ratio and the distance to 
its nearest neighbor, $D_{NN}$.  We use the main galaxy sample of SDSS~DR7 to 
find the potential neighbors.  When calculating the distance to the nearest 
neighbor, we assume a flat geometry and use the linear approximation of Hubble's 
law to define the radial distance from the Earth.   The relationship between 
both \MratioS and \Mratio and $D_{NN}$ is shown in 
Figure~\ref{fig:MtMv_DNN_All}, where we observe an increase in the mass ratio 
with the distance from the nearest neighbor.  This agrees with the simulation 
predictions of \cite{Martizzi20}, where galaxies in less dense environments are 
expected to have less stellar mass for a given dark matter halo mass.  We 
surmise that this is because isolated galaxies tend to evolve slower than 
galaxies in denser regions, as isolated galaxies have lower probabilities of 
star formation episodes resulting from galaxy-galaxy interactions.

Figure~\ref{fig:MtMv_DNN_All} also shows that the relationship between the mass 
ratio and galaxy separation does not change with the inclusion of \MHI.  This 
further supports that the decreasing \Mratio with decreasing distance is a 
result of an increase in stellar mass, a sign that these galaxies have 
experienced more star formation as a result of galaxy-galaxy interactions.

\section{Conclusions}\label{sec:conclusions}
We investigate the possible correlation between a galaxy's ratio of total to 
stellar mass, \MratioS, and visible mass, \Mratio, and its evolutionary 
population, luminosity, gas-phase metallicity, and local environment.  We 
extract a galaxy's rotation curve using the H$\alpha$ velocity maps from the 
SDSS MaNGA DR15, from which we estimate the galaxy's total mass.   We construct 
a statistical model that well describes the distribution over both \MratioS and 
\Mratio in the observed data while simultaneously evaluating the uncertainty in 
the measurement of each mass ratio.

We find that galaxies with the highest values of both \MratioS and \Mratio are 
in the blue cloud, are faint, and have low metallicities.  Based on their 
luminosity and metallicity, these galaxies must be relatively early in their 
star formation histories.  The addition of neutral hydrogen significantly 
reduces the value of \Mratio relative to \MratioS for these galaxies, implying 
that the feedback mechanism at the low-mass end of the stellar-halo mass 
relation is not as extreme as it appears.

We also observe an increase in both \MratioS and \Mratio for the brightest 
galaxies with high metallicities, typically members of the red sequence and 
green valley.  The addition of neutral hydrogen has a smaller effect on \Mratio 
relative to \MratioS for this galaxy population compared to the faint, low 
metallicity galaxies.  Due to their high metallicity, membership in the red 
sequence, and location in the BPT diagram, the star formation processes in these 
galaxies are likely to have been quenched by AGN activity.  This AGN activity is 
also likely the cause for the increase in both \MratioS and \Mratio relative to 
intermediate-mass galaxies, having a similar impact as supernova feedback on 
low-mass galaxies.

We find that both \MratioS and \Mratio increase slightly with the distance to 
the nearest neighbor.  This supports the idea that galaxy interactions have a 
tendency to increase the total visible mass of a galaxy.  

Finally, we derive a parametrization with which we can predict \MratioS based on 
easily accessible galaxy characteristics: the absolute magnitude and 
evolutionary classification.  This empirical relationship can be applied in LSS 
studies to any galaxy with a known redshift and  photometry.

\section*{Acknowledgements}

The authors would like to thank Satya Gontcho A Gontcho for useful discussions 
and insightful questions, Eric Blackman for careful reading and thoughtful 
comments, and the anonymous referees for their detailed comments and 
suggestions.  R.D. acknowledges support from the Department of Energy under the 
grant DE-SC0008475.0.

This project makes use of the MaNGA-Pipe3D data products.  We thank the IA-UNAM 
MaNGA team for creating this catalogue, and the Conacyt Project CB-285080 for 
supporting them.

Funding for the Sloan Digital Sky Survey IV has been provided by the Alfred P. 
Sloan Foundation, the U.S. Department of Energy Office of Science, and the 
Participating Institutions.  SDSS-IV acknowledges support and resources from the 
Center for High-Performance Computing at the University of Utah.  The SDSS web 
site is www.sdss.org.

SDSS-IV is managed by the Astrophysical Research Consortium for the 
Participating Institutions of the SDSS Collaboration including the Brazilian 
Participation Group, the Carnegie Institution for Science, Carnegie Mellon 
University, the Chilean Participation Group, the French Participation Group, 
Harvard-Smithsonian Center for Astrophysics, Instituto de Astrof\'isica de 
Canarias, The Johns Hopkins University, Kavli Institute for the Physics and 
Mathematics of the Universe (IPMU) / University of Tokyo, the Korean 
Participation Group, Lawrence Berkeley National Laboratory, Leibniz Institut 
f\"ur Astrophysik Potsdam (AIP),  Max-Planck-Institut f\"ur Astronomie (MPIA 
Heidelberg), Max-Planck-Institut f\"ur Astrophysik (MPA Garching), 
Max-Planck-Institut f\"ur Extraterrestrische Physik (MPE), National Astronomical 
Observatories of China, New Mexico State University, New York University, 
University of Notre Dame, Observat\'ario Nacional / MCTI, The Ohio State 
University, Pennsylvania State University, Shanghai Astronomical Observatory, 
United Kingdom Participation Group, Universidad Nacional Aut\'onoma de M\'exico, 
University of Arizona, University of Colorado Boulder, University of Oxford, 
University of Portsmouth, University of Utah, University of Virginia, University 
of Washington, University of Wisconsin, Vanderbilt University, and Yale 
University.

\bibliographystyle{aasjournal}
\bibliography{Doug1021_sources}

\end{document}